\documentclass[apjl,twocolumn]{openjournal}
\usepackage{natbib} 
\usepackage[dvipsnames]{xcolor}
\usepackage{aas_macros} 
\usepackage{amssymb}
\usepackage{amsmath}
\usepackage{threeparttable}
\usepackage{multirow}
\usepackage[title]{appendix}
\usepackage{hyperref}	% Hyperlinks
\hypersetup{colorlinks=true,linkcolor=blue,citecolor=blue,filecolor=blue,urlcolor=blue}
\usepackage[caption=false]{subfig}

\newcommand{\cii}{[C\,{\sc ii}]}
\newcommand{\ciiiopt}{C\,{\sc iii}]}
\newcommand{\civopt}{C\,{\sc iv}}
\newcommand{\oiii}{[O\,{\sc iii}]$_{88}$}
\newcommand{\oiiinowave}{[O\,{\sc iii}]}
\newcommand{\oiiidopt}{[O\,{\sc iii}]$_{4959,5007}$}
\newcommand{\oiiiopt}{[O\,{\sc iii}]$_{5007}$}

\newcommand{\hii}{H\,{\sc ii}}

\newcommand{\oiiil}{[O\,{\sc iii}] 88\,$\mu{\rm m}$}
\newcommand{\ciil}{[C\,{\sc ii}] 158\,$\mu{\rm m}$}

\begin{document}

\title[O$^{++}$ at $z>8$]{A first systematic study of \oiiil{} at $\MakeLowercase{z}>8$: two luminous oxygen lines and a powerful ionized outflow in the first 600 million years \vspace{-15mm}}

\author{Hiddo S. B. Algera$^{1\,*}$}
\author{John R. Weaver$^{2}$}
\author{Tom J. L. C. Bakx$^{3}$}
\author{Manuel Aravena$^{4,5}$}
\author{Rychard J. Bouwens$^{6}$}
\author{Karin Cescon$^{6}$}
\author{Chian-Chou Chen$^{1}$}
\author{Elisabete da Cunha$^{7}$}
\author{Pratika Dayal$^{8,9,10}$}
\author{Andreas Faisst$^{11}$}
\author{Andrea Ferrara$^{12}$}
\author{Seiji Fujimoto$^{9,13}$}
\author{Takuya Hashimoto$^{14}$}
\author{Kasper Heintz$^{15,16,17}$}
\author{Rodrigo Herrera-Camus$^{18,5}$}
\author{Jacqueline Hodge$^{6}$}
\author{Hanae Inami$^{19}$}
\author{Akio K. Inoue$^{20,21}$}
\author{Jorryt Matthee$^{22}$}
\author{Romain Meyer$^{17}$}
\author{Shoichiro Mizukoshi$^{1}$}
\author{Chayan Mondal$^{23}$}
\author{Themiya Nanayakkara$^{24}$}
\author{Pascal A. Oesch$^{17,15,16}$}
\author{Andrea Pallottini$^{12,25}$}
\author{Huub R\"{o}ttgering$^{6}$}
\author{Lucie E. Rowland$^{6}$}
\author{Sander Schouws$^{6}$}
\author{Renske Smit$^{26}$}
\author{Laura Sommovigo$^{27}$}
\author{Daniel P. Stark$^{28}$}
\author{Yuma Sugahara$^{20,21}$}
\author{Livia Vallini$^{29}$}
\author{Bovornpratch Vijarnwannaluk$^{1}$}
\author{Paul van der Werf$^{6}$}
\author{Norbert Werner$^{30}$}
\author{Joris Witstok$^{15,16}$}
\author{Mengyuan Xiao$^{17}$
\vspace*{0.3cm}}

\thanks{$^*$E-mail: \href{mailto:hsbalgera@asiaa.sinica.edu.tw}{hsbalgera@asiaa.sinica.edu.tw}}

\affiliation{$^{1}$Institute of Astronomy and Astrophysics, Academia Sinica, 11F of Astronomy-Mathematics Building, No.1, Sec. 4, Roosevelt Rd, Taipei 106319, Taiwan, R.O.C.}
\affiliation{$^{2}$MIT Kavli Institute for Astrophysics and Space Research, Massachusetts Institute of Technology, Cambridge, MA 02139, USA}
\affiliation{$^{3}$Department of Space, Earth and Environment, Chalmers University of Technology, SE-412 96 Gothenburg, Sweden}
\affiliation{$^{4}$Instituto de Estudios Astrof\'{\i}cos, Facultad de Ingenier\'{\i}a y Ciencias, Universidad Diego Portales, Av. Ej\'ercito 441, Santiago, Chile}
\affiliation{$^{5}$Millenium Nucleus for Galaxies (MINGAL)}
\affiliation{$^{6}$Leiden Observatory, Leiden University, P.O. Box 9513, 2300 RA Leiden, The Netherlands}
\affiliation{$^{7}$International Centre for Radio Astronomy Research, University of Western Australia, 35 Stirling Hwy., Crawley, WA 6009, Australia}
\affiliation{$^{8}$Canadian Institute for Theoretical Astrophysics, 60 St George St, University of Toronto, Toronto, ON M5S 3H8, Canada}
\affiliation{$^{9}$David A. Dunlap Department of Astronomy and Astrophysics, University of Toronto, 50 St George St, Toronto ON M5S 3H4, Canada}
\affiliation{$^{10}$Department of Physics, 60 St George St, University of Toronto, Toronto, ON M5S 3H8, Canada}
\affiliation{$^{11}$
IPAC, California Institute of Technology, 1200 E. California Blvd., Pasadena, CA 91125, USA}
\affiliation{$^{12}$Scuola Normale Superiore, Piazza dei Cavalieri 7, I-56126 Pisa, Italy}
\affiliation{$^{13}$Dunlap Institute for Astronomy and Astrophysics, 50 St. George Street, Toronto, Ontario, M5S 3H4, Canada}
\affiliation{$^{14}$Division of Physics, Faculty of Pure and Applied Sciences, University of Tsukuba, Tsukuba, Ibaraki 305-8571, Japan}
\affiliation{$^{15}$Cosmic Dawn Center (DAWN), Copenhagen, Denmark}
\affiliation{$^{16}$Niels Bohr Institute, University of Copenhagen, Jagtvej 155A, Copenhagen N, DK-2200, Denmark}
\affiliation{$^{17}$Department of Astronomy, University of Geneva, Chemin Pegasi 51, 1290 Versoix, Switzerland}
\affiliation{$^{18}$Departamento de Astronomía, Universidad de Concepción, Barrio
Universitario, Concepción, Chile}
\affiliation{$^{19}$Hiroshima Astrophysical Science Center, Hiroshima University, 1-3-1 Kagamiyama, Higashi-Hiroshima, Hiroshima 739-8526, Japan}
\affiliation{$^{20}$Department of Physics, School of Advanced Science and Engineering, Faculty of Science and Engineering, Waseda University, 3-4-1 Okubo, Shinjuku, Tokyo 169-8555, Japan}
\affiliation{$^{21}$Waseda Research Institute for Science and Engineering, Faculty of Science and Engineering, Waseda University, 3-4-1 Okubo, Shinjuku, Tokyo 169-8555, Japan}
\affiliation{$^{22}$Institute of Science and Technology Austria (ISTA), Am Campus 1, 3400 Klosterneuburg, Austria}
\affiliation{$^{23}$S. N. Bose National Centre for Basic Sciences Block-JD, Sector-III, Salt Lake, Kolkata-700106, India}
\affiliation{$^{24}$Centre for Astrophysics and Supercomputing, Swinburne University of Technology, P.O. Box 218, Hawthorn, 3122, VIC, Australia}
\affiliation{$^{25}$Dipartimento di Fisica “Enrico Fermi”, Universitá di Pisa, Largo Bruno Pontecorvo 3, Pisa I-56127, Italy}
\affiliation{$^{26}$Astrophysics Research Institute, Liverpool John Moores University, 146 Brownlow Hill, Liverpool L3 5RF, UK}
\affiliation{$^{27}$Center for Computational Astrophysics, Flatiron Institute, 162 5th Avenue, New York, NY 10010, USA}
\affiliation{$^{28}$Department of Astronomy, University of California, Berkeley, Berkeley, CA 94720, USA}
\affiliation{$^{29}$INAF -- Osservatorio di Astrofisica e Scienza dello Spazio di Bologna, Via Gobetti 93/3, 40129 Bologna, Italy}
\affiliation{$^{30}$Department of Theoretical Physics and Astrophysics, Masaryk University, Brno 61137, Czechia}

\begin{abstract}
We present deep ALMA Band 7 observations of the \oiiil{} line and underlying dust continuum emission in four UV-bright, gravitationally lensed (magnification $\mu = 1.4-3.8$), \textit{JWST}-selected galaxies at $z = 8.5 - 10.3$, with observed magnitudes $-22.5 \lesssim M_\mathrm{UV} \lesssim -20.5$. \oiiil{} is confidently detected in UNCOVER-10646 at $z=8.5080 \pm 0.0011$ ($15\sigma$) and DHZ1 at $z=9.3113 \pm 0.0006$ ($6\sigma$), with both being intrinsically luminous systems [$L_\text{\oiiinowave{}} = (1.1 - 1.6) \times 10^9\,L_\odot$] that follow the local \oiiinowave{}-SFR relation. \oiiil{} remains undetected in the two $z>10$ targets, including in the $z=10.07$ X-ray AGN UHZ1, where we obtain a deep limit of $L_\text{\oiiinowave{}} < 6 \times 10^7\,L_\odot$. Dust emission is not detected in any individual source nor in a stack ($<3\sigma$). The high S/N \oiiil{} detection in UNCOVER-10646 uniquely reveals an additional broad component ($\mathrm{FWHM} = 1366_{-329}^{+473}\,\mathrm{km/s}$; $\Delta\mathrm{BIC}\approx20$) indicative of an ionized outflow. We infer a high outflow rate of $\dot{M}_\mathrm{out} = 128_{-46}^{+80}\,M_\odot\,\mathrm{yr}^{-1}$, corresponding to a mass loading factor $\eta = \dot{M}_\mathrm{out}/\mathrm{SFR} = 2.9_{-1.0}^{+1.8}$ that matches or exceeds theoretical predictions and \textit{JWST}-based studies of ionized outflows at high redshift. While high-resolution ALMA follow-up is required to confirm and spatially resolve the outflow, this first systematic study at $z>8$ highlights the unique diagnostic power of \oiiil{} in characterizing galaxies in the early Universe. 
\end{abstract}

\section{Introduction}
\label{sec:introduction}

In the last few years, observations from the \textit{James Webb Space Telescope} (\textit{JWST}) have pushed the observational frontier of galaxy formation well into the first few hundred Myr after the Big Bang \citep[e.g.,][]{arrabal-haro2023,curtis-lake2023,carniani2024,harikane2024,naidu2025,roberts-borsani2025}. One of the most puzzling yet exciting findings is that the number densities of UV-luminous systems at $z>10$ exceed predictions from pre-\textit{JWST} galaxy formation theory \citep[e.g.,][]{bouwens2023,finkelstein2023,harikane2023,adams2024,chemerynska2024_uvlf,donnan2024,whitler2025}. Accordingly, understanding the formation, evolution and eventual fate of these early `over-luminous' galaxies is one of the key objectives within current high-redshift astronomy.

Prior to the advent of the \textit{JWST}, one of the prime ways of spectroscopically confirming galaxies in the Epoch of Reionization (EoR; $z\gtrsim6$) was through the Lyman-$\alpha$ line \citep[e.g.,][]{zitrin2015,roberts-borsani2016,stark2017}. However, this line is easily attenuated by neutral hydrogen, both within the interstellar medium (ISM) in and circumgalactic medium (CGM) around the galaxy itself, and -- at the highest redshifts -- by an increasingly neutral intergalactic medium (IGM) \citep[e.g.,][]{mason2018,mason2019,bosman2022}. Instead, it was realized that the Atacama Large Millimeter/submillimeter Array (ALMA) could efficiently spectroscopically confirm galaxies through bright far-infrared (FIR) fine structure lines \citep[e.g.,][]{ouchi2013,inoue2014}, which have the additional benefit that they are not strongly affected by dust attenuation. One of the brightest such FIR lines is the \ciil{} emission line (hereafter \cii{}), which has since become the primary means of studying $z\gtrsim4$ galaxies at far-infrared wavelengths \citep{capak2015,smit2018,lefevre2020,bethermin2020,faisst2020,bouwens2022,schouws2023,herrera-camus2025}. Not only does a \cii{} detection with ALMA yield a precise spectroscopic redshift, it simultaneously provides information on the star formation rate \citep[SFR; e.g.,][]{delooze2014,herrera-camus2018a,schaerer2020}, ISM contents \citep[e.g.,][]{zanella2018,dessauges-zavadsky2020,heintz2022,aravena2023}, and kinematics of the galaxy \citep[e.g.,][]{jones2021,rowland2024,fujimoto2025_grapes}. 

However, in the distant galaxy population ($z\gtrsim6$), the \oiiil{} line (hereafter \oiii{}) frequently appears brighter than \cii{} \citep[e.g.,][though see \citealt{marrone2018,walter2018,hashimoto2019_quasar,bakx2024}]{inoue2016,hashimoto2019,bakx2020,carniani2020,harikane2020,witstok2022,ren2023,algera2024,algera2025_oiii,schouws2025}, making it a more efficient line for spectroscopically confirming and characterizing the most distant galaxies \citep[e.g.,][]{inoue2014,inoue2016,bouwens2022}. This is likely due to most high-redshift galaxies being characterized by young, bursty star formation, with strong radiation fields and/or low metallicities. As highlighted by several theoretical works \citep[e.g.,][]{katz2017,katz2019,olsen2017,vallini2017,vallini2021,vallini2024,pallottini2019,pallottini2022,arata2020,kohandel2025,nakazato2025}, these conditions preferentially boost the luminosity of the \oiii{} line, while potentially simultaneously suppressing that of \cii{}.

The power of ALMA in characterizing the most distant galaxies was already demonstrated by the \oiii{} detections of two $z>8$ galaxies in the pre-\textit{JWST} era -- MACS0416-Y1 at $z=8.31$ \citep{tamura2019,tamura2023} and MACS1149-JD1 at $z=9.11$ \citep{hashimoto2018,tokuoka2022}, with the latter being the highest redshift robust emission line detection until the advent of the \textit{JWST}. Now that the \textit{JWST} has been operational for a few years, several newly-identified high-redshift sources have been followed up by ALMA. \citet{schouws2024} and \citet{carniani2024_oiii} simultaneously and independently reported an \oiii{} detection from the $z=14.18$ galaxy JADES-GS-z14-0, which was previously spectroscopically confirmed through a Lyman break detection by \citet[][see also \citealt{helton2024}]{carniani2024}. Further \oiii{} detections of $z>8$ galaxies were reported by \citet[][$z=8.49$]{fujimoto2024_s4590}, \citet[][$z=11.12$]{witstok2025}, and \citet[][$z=12.33$]{zavala2024_alma}, emphasizing the tremendous power of ALMA in detecting even the most distant systems.

One of the prime strengths of ALMA is its high velocity resolution, which -- being tuneable -- is typically within the range of $20-50\,\mathrm{km/s}$ for high-redshift galaxy studies. This enables determining spectroscopic redshifts with a high accuracy ($\Delta z \lesssim 10^{-3}$), especially compared to redshifts obtained from the Lyman break with \textit{JWST}/NIRSpec prism spectroscopy \citep[e.g.,][]{schouws2024,carniani2024_oiii}. The latter can be uncertain by well beyond $\Delta z > 0.2$, and moreover appear to be systematically biased high due to Damped Lyman-$\alpha$ Absorption (DLA) by neutral gas around the galaxy \citep[e.g.,][]{fujimoto2023,deugenio2024,hainline2024,asada2025,heintz2025_primal,rowland2025_dla,witstok2025}. Moreover, at $z\gtrsim10$, the brightest rest-optical lines such as \oiiiopt{} shift out of the NIRSpec coverage, thus requiring one to either shift to the less sensitive MIRI instrument \citep[e.g.,][]{hsiao2024,zavala2025_miri}, rely on typically fainter rest-UV lines such as \ciiiopt{} and \civopt{} \citep[e.g.,][]{bunker2023,naidu2025}, or turn to ALMA for spectroscopic confirmation and characterization \citep[e.g.,][]{carniani2024_oiii,schouws2024,schouws2025,witstok2025}.

Even compared to the high-resolution gratings available with NIRSpec, ALMA's velocity resolution exceeds that of \textit{JWST}, facilitating detailed kinematic studies of reionization-era galaxies, potentially revealing merging activity, rotating disks, and outflows \citep[e.g.,][]{akins2022,tokuoka2022,rowland2024,knudsen2025,scholtz2025_gsz14}. In principle, ALMA is also able to attain a higher spatial resolution by combining its longest baselines and high-frequency receivers, although this requires sufficiently bright galaxies to fully exploit \citep[e.g.,][]{ikeda2025_fossils,walter2025}.

It is in particular the combination of ALMA and \textit{JWST}, however, that enables unique constraints on the ISM properties of distant galaxies. For instance, far-infrared-to-rest-optical line ratios can be used to constrain metallicities, densities, and ionization parameters \citep[e.g.,][]{heintz2023,carniani2024_oiii,fujimoto2024_s4590,algera2025_oiii,schouws2024}, although some uncertainties regarding the precise interpretation of rest-UV/optical and FIR line ratios remain \citep[e.g.,][]{harikane2025,usui2025}. Furthermore, ALMA and \textit{JWST} enable detailed studies of the build-up of dust in distant galaxies \citep[e.g.,][]{heintz2023,heintz2025_zD1,algera2025_hz10,algera2026,ciesla2025,mitsuhashi2025,umehata2025}. Dust is now known to be widespread in massive galaxies and quasars at $z\approx6-8$ \citep[e.g.,][]{bowler2018,marrone2018,venemans2018,venemans2020,hashimoto2019,tamura2019,bakx2021,bakx2025,decarli2022,inami2022,schouws2022,witstok2022,algera2023,algera2024,algera2026,barrufet2023,tripodi2024}, although its importance and ubiquity at even earlier epochs remains unknown. Given the ability of the \textit{JWST} to study dust through attenuation \citep[e.g.,][]{witstok2023_2175,fisher2025,markov2025,shivaei2025}, this enables a powerful -- albeit still somewhat under-exploited -- synergy with ALMA-based studies of dust in emission.

In this work, we perform the first systematic study of the \oiiil{} emission line in a sample of $z>8$ galaxies. We present new, sensitive ALMA \oiii{} and continuum observations towards four galaxies at $z=8.5 - 10.3$, as detailed in Section \ref{sec:observations}. We present our results in Section \ref{sec:results}, and discuss the \oiii{}-star formation rate relation as well as a possible ionized outflow signature in our brightest target in Section \ref{sec:discussion}. Finally, we summarize our conclusions in Section \ref{sec:conclusions}. Throughout this work, we adopt a standard $\Lambda$CDM cosmology with $H_0=70\,\text{km\,s}^{-1}\text{\,Mpc}^{-1}$, $\Omega_m=0.30$ and $\Omega_\Lambda=0.70$, and a solar oxygen abundance of $12 + \log(\mathrm{O/H}) = 8.69$, following \citet{asplund2009}.

\section{Observations \& Methods}
\label{sec:observations}

\subsection{Target selection}
\label{sec:observations_targets}

\begin{figure}
    \centering
    \includegraphics[width=0.5\textwidth]{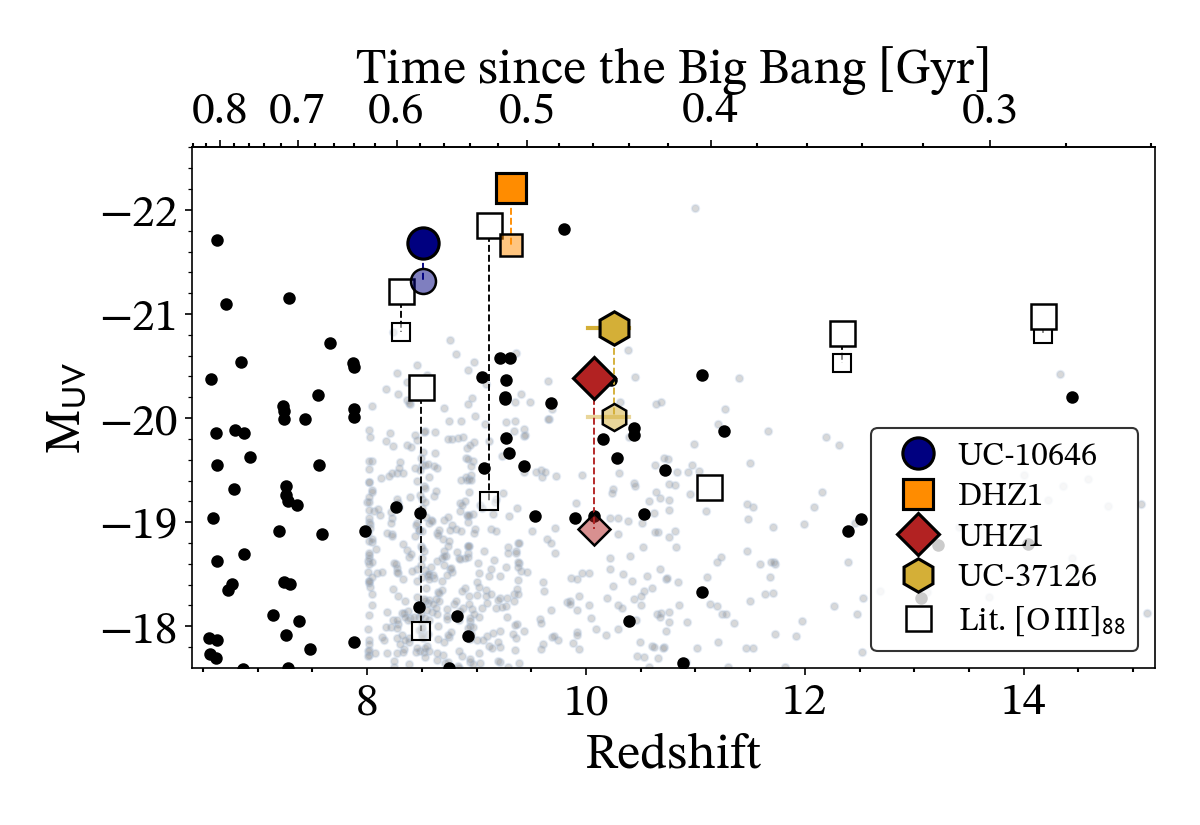}
    \caption{Absolute UV magnitude versus redshift (bottom $x$-axis) and Cosmic time (top axis) for our targets (large, colored markers) and other $z>8$ galaxies targeted in \oiii{} in the literature \citep[white squares;][]{hashimoto2018,tamura2019,fujimoto2024_s4590,zavala2024_alma,carniani2024_oiii,schouws2024,witstok2025}. The larger, opaque markers show the apparent brightness, including the effects of gravitational lensing, while the smaller symbols -- connected via the dashed lines -- correspond to de-lensed values. We moreover show the compilation of $z>6.5$ galaxies spectroscopically confirmed with the \textit{JWST} from \citet[][solid black markers]{tang2024,tang2025} as well as the $z>8$ photometric candidates from \citet[][semi-transparent gray markers]{hainline2024}. Our targets are particularly UV-luminous galaxies at $z>8$.}
    \label{fig:Muv_redshift}
\end{figure}

\begin{table*}
    \def\arraystretch{1.4}
    %\addtolength{\tabcolsep}{-0.5pt} % slightly reduce column spacing to make the table fit
    \centering
    \caption{Physical properties of our four targets}
    \label{tab:properties}
    \begin{threeparttable}
    \begin{tabular}{lccccccccc}
    
    \hline\hline
    ID & RA & Dec & $z_\mathrm{JWST}$ & $z_\text{\oiii{}}$ & $M_\mathrm{UV}$ & $\mu$ & $\mathrm{SFR}/(M_\odot\,\mathrm{yr}^{-1})$ & $\log(M_\star / M_\odot)$ & Refs \\
    \hline 
    
UNCOVER-10646 & 3.636975 & -30.406361 & $8.513 \pm 0.001$ & $8.5080 \pm 0.0011$ & $-21.34 \pm 0.04$ & $1.40 \pm 0.01$ & $45.4 \pm 1.7 $& $8.62 \pm 0.05$ & 1,2 \\
DHZ1 & 3.617193 & -30.425535 & $9.3127 \pm 0.0002$ & $9.3113 \pm 0.0006$ & $-21.66 \pm 0.03$ & $1.66 \pm 0.02$ & $19_{-6}^{+5}$ & $9.2_{-0.2}^{+0.1}$ & 3 \\
UHZ1 & 3.567071 & -30.377861 & $10.073 \pm 0.002$ & $-$ & $-18.93 \pm 0.10$ & $3.81_{-0.56}^{+0.41}$ & $1.25_{-0.12}^{+0.18}$ & $8.14_{-0.11}^{+0.09}$ & 4,5 \\
UNCOVER-37126 & 3.590111 & -30.359742 & $10.02 - 10.39$ & $-$ & $-20.01 \pm 0.10$ & $2.20 \pm 0.01$ & $4.57_{-1.43}^{+0.85}$ & $8.63 \pm 0.06$ & 1,6 \\

\hline\hline 
\end{tabular}
    \textit{Notes}: Col.\ (1): Galaxy identifier. Col.\ (2,3) Right ascension and declination, in degrees. Col.\ (4) Spectroscopic redshift from \textit{JWST}/NIRSpec. Col.\ (5) Spectroscopic redshift obtained from the \oiiil{} line in this work, where detected. Col.\ (6) Intrinsic (i.e., lensing-corrected) UV magnitude (AB). Col.\ (7) Gravitational lensing magnification. Col.\ (8) Lensing-corrected star formation rate. Col.\ (9) Lensing-corrected stellar mass. Col.\ (10) List of references: \citet[][1]{fujimoto2024_uncover}; Weaver et al.\ (in preparation, 2); \citet[][3]{boyett2024}; \citet[][4]{goulding2023}; \citet[][5]{bogdan2024}; \citet[][6]{wang2024_uncover_catalog}.
\end{threeparttable}
\end{table*}

As part of the ALMA Cycle 11 project 2024.1.00440.S (PI Algera), we obtained Band 7 \citep{mahieu_band7} observations towards four high-redshift galaxies spanning $z=8.5 - 10.3$. These galaxies were selected to be UV-luminous (apparent $M_\mathrm{UV} \lesssim -21$; not corrected for lensing magnification -- see below) with prior spectroscopic redshifts from \textit{JWST}/NIRSpec (either from emission lines or the Lyman break) placing them at $8 < z < 11$, based on catalogs publicly available by the ALMA Cycle 11 proposal deadline in April 2024 \citep[e.g.,][]{tang2024,heintz2025_primal}. Moreover, to facilitate the ALMA observations, the targets were required to lie at southern or equatorial declinations ($\delta < 20^\circ$). This selection yielded three galaxies, all of which were targeted in our ALMA program. The targets are found behind the Abell 2744 lensing cluster observed as part of the \textit{JWST} UNCOVER survey \citep{bezanson2024}, and includes two bright galaxies first reported by \citet{fujimoto2024_uncover} at $z=8.51$ and $z\approx10.3$ (UNCOVER IDs 10646 and 37126, respectively) whose properties are discussed in further detail below. A third UV-luminous galaxy at $z=9.31$, first reported by \citet{boyett2024} and known both as DHZ1 and Gz9p3, was also selected. Although not formally making our UV magnitude cut, we also targeted UHZ1 at $z=10.07$, an X-ray AGN also in the Abell 2744 field reported by \citet{bogdan2024} and \citet{goulding2023}, completing the sample of four. This galaxy was reported to have a $\sim2\sigma$ dust continuum signal in the wide-field but shallow ALMA observations  from \citet{fujimoto2025_dualz}, motivating our deeper continuum and simultaneous \oiii{} follow-up. The \oiii{} line and underlying continuum of UHZ1 were also separately targeted in a different Cycle 11 program (2024.1.01197.S, PI Fujimoto), and we include these observations here for completeness. 

Our targets are shown in the context of the high-redshift galaxy population identified with the \textit{JWST} in Figure \ref{fig:Muv_redshift}, and we briefly discuss them individually in what follows. 

\subsubsection{UNCOVER-10646}
\label{sec:UNCOVER_10646}

UNCOVER-10646 was identified as a UV-luminous galaxy ($M_\mathrm{UV} = -21.3$) at $z=8.511 \pm 0.001$ by \citet{fujimoto2024_uncover} based on the detection of a clear Lyman break and several emission lines in NIRSpec multi-shutter array (MSA) prism spectroscopy. The galaxy is magnified by a modest factor of $\mu = 1.40 \pm 0.01$ based on the lensing models by \citet{furtak2023_lensing}. UNCOVER-10646 is included in the sample of extreme UV-line emitters from \citet{treiber2025}, who identify it as a likely merger of two components, with the source appearing resolved in blue NIRCam imaging, but blended in the redder F444W band. As detailed in Weaver et al.\ (in preparation), the galaxy pair can be spectrally separated in the NIRSpec prism observations, which reveals that (at least) one of the two components is likely an AGN based on its strong UV lines. Given the close separation of the pair, however, it is not resolved in our \oiii{} observations (see below). Weaver et al.\ (in preparation) suggest that the system is likely a late-stage major merger, and in this work we therefore treat it as a singular system, but comment on its merger nature where relevant. Weaver et al.\ moreover fit the integrated NIRSpec spectrum and NIRCam photometry of UNCOVER-10646 with {\sc{Bagpipes}} \citep{carnall2018,carnall2019} to infer its physical properties, assuming a non-parametric star-formation history (SFH) and a \citet{calzetti2000} dust attenuation curve. The stellar continuum of UNCOVER-10646 appears galaxy-dominated, and consequently no AGN components are included in the fit. The adopted stellar mass and SFR from Weaver et al.\ (in preparation) are presented in Table \ref{tab:properties}. We moreover note that the adopted SFR agrees with that inferred directly from the H$\beta$ line.

\subsubsection{DHZ1 / Gz9p3}
\label{sec:DHZ1}

DHZ1, also known as Gz9p3, is a UV-luminous galaxy at $z=9.3127 \pm 0.0002$. It was first spectroscopically confirmed by \citet{boyett2024} through high-resolution ($R\sim2700$) \textit{JWST}/NIRSpec spectroscopy using the F100LP/G140H, F170LP/G235H and F290LP/G395H filter/grating combinations. The galaxy is gravitationally lensed by $\mu = 1.66 \pm 0.02$ \citep{furtak2023_lensing,boyett2024} and appears spatially extended in NIRCam imaging, likely due to an ongoing interaction. \citet{boyett2024} determine the physical properties of DHZ1 through SED-fitting with {\sc{Bagpipes}}, both for the core of the galaxy which is covered by the NIRSpec slit, and for the full system through fitting only the photometry. We here adopt the stellar mass and SFR inferred for the complete system (Table \ref{tab:properties}), as our \oiii{} observations are only marginally spatially resolved, and are therefore sensitive to emission from the entire galaxy. The SFR measured by \citet{boyett2024} corresponds to the value averaged over the last $100\,\mathrm{Myr}$ as obtained from {\sc{Bagpipes}}, where a log-normal star formation history (SFH) and a \citet{calzetti2000} dust attenuation law were assumed. A consistent global SFR was recently inferred by \citet{bik2026} using the H$\alpha$ line, which traces star formation on shorter ($\sim10\,\mathrm{Myr}$) timescales.

\subsubsection{UHZ1}
\label{sec:UHZ1}

UHZ1 was first reported as a compact high-redshift galaxy candidate by \citet{atek2023} and \citet{castellano2023}, and was subsequently found to be X-ray-detected in deep \textit{Chandra} imaging of the Abell 2744 field by \citet{bogdan2024}, implying the galaxy hosts a luminous AGN. Its high-redshift nature of $z=10.073 \pm 0.002$ was subsequently confirmed through the NIRSpec prism detection of several emission lines by \citet{goulding2023}. UHZ1 is gravitationally lensed by a factor of $\mu\approx4$, and we adopt the lensing-corrected physical properties of the galaxy obtained by \citet[][Table \ref{tab:properties}]{goulding2023}. They perform {\sc{Bagpipes}} fitting to its NIRCam photometry and NIRSpec spectrum using a delayed-$\tau$ star formation history and a \citet{charlot_fall2000} dust model. No AGN component is included in the fit, as \citet{goulding2023} find no clear evidence of the AGN contributing to the rest-frame UV emission of the galaxy, given that no strong UV lines are detected.

Recently, MIRI/LRS (Low-Resolution Spectrograph) observations of UHZ1 were carried out by \citet{alvarez-marquez2026}, who detected the \oiiidopt{}+H$\beta$ complex as well as the H$\alpha$ line in the galaxy. These observations suggest a slightly lower redshift of $z=10.054 \pm 0.011$ for UHZ1, which is offset from the NIRSpec-based value from \citet{goulding2023} by $\sim460\,\mathrm{km/s}$. Both the NIRSpec and MIRI observations were conducted at low spectral resolution, and in this work we adopt the NIRSpec-based redshift while we discuss the alternative MIRI redshift in Appendix \ref{app:spectra1D_nondet}. We moreover note that the SFR and stellar mass of UHZ1 inferred by \citet{goulding2023} and \citet{alvarez-marquez2026} are fully consistent with one another.

\subsubsection{UNCOVER-37126}
\label{sec:UNCOVER-37126}

UNCOVER-37126 is a UV-luminous galaxy ($M_\mathrm{UV} = -20.01$) first reported by \citet{fujimoto2024_uncover}, who inferred a spectroscopic redshift of $z=10.255$ based on a NIRSpec prism detection of the Lyman break. As discussed in several works, Lyman-break-derived redshifts can be uncertain by $\Delta z > 0.2$ \citep[e.g.,][]{hainline2024,heintz2025_primal,witstok2025}, and subsequent studies have indeed reported different values for UNCOVER-37126, ranging from $z=10.02 - 10.39$ \citep{tang2024,heintz2025_primal,price2025,roberts-borsani2025}. We here adopt the redshift of $z=10.255$ from \citet{fujimoto2024_uncover} as fiducial, but discuss the effects of its uncertain redshift in the context of our \oiii{} observations in the subsequent sections. We furthermore note that MIRI/LRS observations of UNCOVER-37126, as recently presented by \citet{marques-chaves2026}, did not detect any rest-optical emission lines in the galaxy, and hence cannot be used to refine the redshift measurement.

We adopt the physical properties and lensing magnification of UNCOVER-37126 from the UNCOVER DR4-spec catalog, where spectroscopic redshifts and medium-band photometry have been incorporated into the SED fitting \citep{furtak2023_lensing,suess2024_megascience,wang2024_uncover_catalog,weaver2024_uncover_catalog,price2025}. Fitting was performed using {\sc{Prospector}} \citep{johnson2021} under the assumption of a non-parametric star formation history. A two-component dust model is adopted following \citet{charlot_fall2000}, with a flexible attenuation curve as in \citet{noll2009}. For the SFR of UNCOVER-37126, we adopt the value averaged over the last $30\,\mathrm{Myr}$ (Table \ref{tab:properties}). \\

We acknowledge that the SED fitting approaches employed to determine the stellar masses and SFRs of our targets differ slightly from one another. Given that we do not perform a detailed comparison of these quantities between the galaxies, we do not attempt to homogenize these measurements at this stage. In a future work, as part of the upcoming Cycle 12 ALMA Large Program PHOENIX (PI Schouws) which will target \oiii{} in 15 additional $z>8$ galaxies, we plan to employ a uniform SED-fitting approach for the complete high-redshift sample observed with ALMA, including the pilot sources presented here. We moreover discuss UV+IR-based star formation rate measurements -- which do not rely on any SED-fitting -- in the context of the \oiii{}-SFR relation in Section \ref{sec:discussion_oiii_sfr}.

\begin{table*}
    \def\arraystretch{1.4}
    %\addtolength{\tabcolsep}{-0.5pt} % slightly reduce column spacing to make the table fit
    \centering
    \caption{Properties of the naturally-weighted ALMA Band 7 data presented in this work.}
    \label{tab:observations}
    \begin{threeparttable}
    \begin{tabular}{lccccccccc}
    
    \hline\hline
    ID & $t_\mathrm{obs}$ & PWV$^a$ & $\nu_\mathrm{cont}$ & $\sigma_\mathrm{RMS,cont}^b$ & $\sigma_\mathrm{RMS,chan}^c$ & $\theta_\mathrm{major}$ & $\theta_\mathrm{minor}$ & $\text{PA}$ & PID \\
    \hline 
    & [min] & [mm] & [GHz] & [$\mu$Jy/beam] & [mJy/beam] & [asec] & [asec] & [deg] & \\
    \hline
    
UNCOVER-10646 & $133.7$ & $0.94 \pm 0.04$ & $349.8$ & $14.4$ & $0.20$ & $0.65$ & $0.54$ & $86.8$ & 2024.1.00440.S \\
DHZ1          & 130.6 & $1.09 \pm 0.05$ & $335.5$ & $18.2$ & $0.34$ & $0.72$ & $0.57$ & $83.3$ & 2024.1.00440.S \\
UHZ1          & 265.4 & $1.08 \pm 0.05$ & $300.0$ & $9.6$ & $0.14$ & $0.80$ & $0.61$ & $-85.5$ & 2024.1.00440.S, 2024.1.01197.S \\
UNCOVER-37126 & 124.9 & $0.77 \pm 0.10$ & $294.6$ & $11.1$ & $0.17$ & $0.34$ & $0.28$ & $80.1$ & 2024.1.00440.S \\

\hline\hline 
\end{tabular}
    \begin{tablenotes}
    \item[a] Mean and standard deviation of the Precipitable Water Vapor (PWV) during the observations, in millimeter.
    \item[b] Continuum RMS noise, based on naturally-weighted imaging that excludes emission-line-contaminated channels for UNCOVER-10646 and DHZ1.
    \item[c] Median RMS noise per $30\,\mathrm{km/s}$ channel across the SPWs in the baseband covering the \oiii{} line.
    \end{tablenotes}
\end{threeparttable}
\end{table*}

\subsection{Observations, calibration and imaging}
\label{sec:observations_imaging}

ALMA observations of our four targets were carried out in Band 7 between 2024 October 02 and 2025 March 19. The total on-source time ranged from $125-265\,\mathrm{min}$, as further detailed in Table \ref{tab:observations}. The visibility data were calibrated using the standard ALMA pipeline, and were restored in {\sc{CASA}} \citep{casa2022} using the {\sc{scriptForPI.py}} retrieved when obtaining the data from the ALMA archive. For UHZ1, we concatenate the data from the two ALMA projects targeting this galaxy -- 2024.1.00440.S and 2024.1.01197.S -- via {\sc{concat}}, and reweight the visibilities via {\sc{statwt}} prior to imaging. No continuum subtraction is applied to any of the four targets, as no dust emission is confidently detected (Section \ref{sec:results_dust}).

We first create naturally-weighted datacubes for our targets using the {\sc{tclean}} task in {\sc{CASA}}. We adopt a channel width of $30\,\mathrm{km/s}$ in the imaging and use the automasking capabilities in {\sc{tclean}} following the recommended parameters for ALMA data \citep{kepley2020}. Cleaning of bright emission in the entire Band 7 field-of-view is performed down to $2\sigma$. 

Subsequently, we produce naturally-weighted continuum images for our targets. As detailed in Section \ref{sec:results}, the \oiii{} line is detected in both UNCOVER-10646 and DHZ1, and we ensure the emission-line-contaminated channels are excluded from the imaging, while for the other two sources all channels are included given that no line is detected. For DHZ1, we simply exclude all channels within $2\times$ the line FWHM obtained in Section \ref{sec:results_oiii}, while the line profile of UNCOVER-10646 appears more complicated than a single Gaussian (Section \ref{sec:discussion_outflow}), leading us to exclude a wider velocity range corresponding to $\pm700\,\mathrm{km/s}$ around the line center to ensure the continuum image is not contaminated by any high-velocity line emission. This effectively excludes $\sim90\%$ of the channels in the spectral window (SPW) centered on the emission line, and we have verified that we obtain consistent results regarding the dust continuum properties of UNCOVER-10646 when this SPW is omitted altogether. To enhance our sensitivity to possible weak, extended dust emission, we also create tapered continuum images spanning a range of beam sizes between the naturally-weighted resolution (Table \ref{tab:observations}) and $1.5''$, in steps of $0.1''$.

\subsection{Line flux measurements}
\label{sec:observations_fluxMeasurements}

To verify whether the \oiii{} line is detected, we first extract a one-dimensional spectrum from the datacubes, adopting circular apertures centered on the rest-UV positions of our targets, and exploring a range of aperture radii between the beam size and $2.0''$. We find that the line is detected in UNCOVER-10646 and DHZ1, while it remains undetected in our two $z>10$ targets. For the two detections, we adopt the iterative approach from \citet{algera2025_oiii} to measure the line flux and FWHM, adopting a fixed aperture radius of $0.8''$ and $1.1''$ for UNCOVER-10646 and DHZ1, respectively, to fully capture any faint extended emission. This approach first creates a moment-0 map around the expected line center, and fits the integrated line emission with a 2D Gaussian. The aperture is then shifted based on the 2D Gaussian centroid if needed, and the flux per channel is extracted from all pixels within the aperture. A 1D Gaussian is fitted to the spectrum, and the moment-0 map is re-extracted across $1.2\times$ the line FWHM.\footnote{A range of $1.2\times$ the FWHM maximizes the signal-to-noise for a Gaussian line \citep[e.g.,][]{novak2020}.} The iterative fitting process is repeated until convergence is reached, and we adopt the area under the final 1D Gaussian fit as the \oiii{} line flux. 

\begin{figure*}
    \centering
    \includegraphics[width=0.95\linewidth]{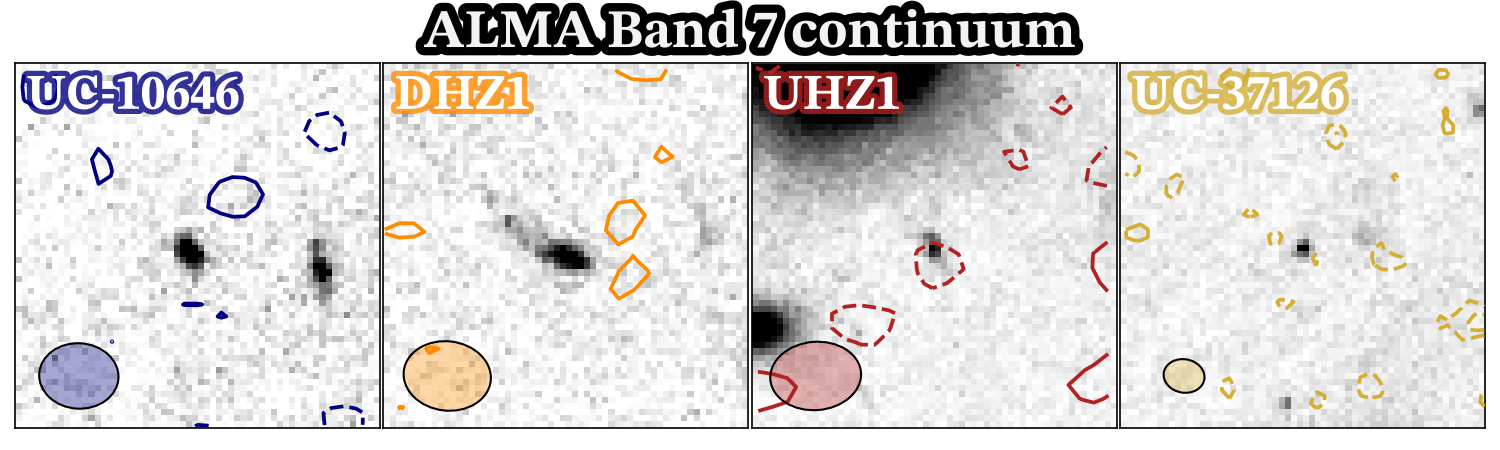}
    \caption{Naturally-weighted ALMA Band 7 dust continuum cutouts ($3''\times3''$) of our four targets, overplotted on \textit{JWST}/NIRCam F150W imaging. Contours are drawn at $\pm2,3,4,\ldots\times\sigma$ intervals, where $\sigma$ is the continuum RMS. Negative contours are dashed. None of the sources are convincingly dust-detected, although UNCOVER-10646 shows a $\sim2\sigma$ continuum signal in tapered imaging (Figure \ref{fig:continuum_UC10646_tapered} in Appendix \ref{app:dust_continuum_UC10646}), tentatively suggesting the presence of an extended dust reservoir.}
    \label{fig:dustContinuum}
\end{figure*}

For UHZ1 and UNCOVER-37126, we adopt $3\sigma$ upper limits on the \oiii{} line flux. We create moment-0 maps across $200\,\mathrm{km/s}$ centered on the NIRSpec-based redshift (i.e., across $\pm100\,\mathrm{km/s}$), and measure the noise $\sigma$ in $0.8''$ and $0.5''$-radius apertures for UHZ1 and UNCOVER-37126, respectively. These apertures are chosen to be slightly larger than the beam size to ensure that any extended \oiii{} emission is accounted for. We note that both galaxies are compact in \textit{JWST}/NIRCam imaging (Figures \ref{fig:dustContinuum} and \ref{fig:momentZero}), such that the presence of significantly extended \oiii{} is unlikely. Moreover, a FWHM of $200\,\mathrm{km/s}$ is conservative in light of recent \oiii{} detections of other (low-mass) $z>8$ galaxies, which typically show narrower linewidths \citep[e.g.,][see also Section \ref{sec:results_oiii}]{zavala2024_alma,carniani2024_oiii,schouws2024,witstok2025}. For UNCOVER-37126, we adopt a fiducial redshift of $z=10.255$, though we caution that the uncertain redshift of this source may have caused the \oiii{} line to be missed, as discussed in Section \ref{sec:observations_targets}. As such, the adopted upper limit is only appropriate if the redshift of UNCOVER-37126 is within the range covered by the ALMA observations. Given our adopted spectral setup, the two sidebands cover a redshift range of $z=10.22 - 10.36$ and $z=10.68 - 10.83$, and we confirm that no line is detected in either sideband (Appendix \ref{app:spectra1D_nondet}).

\section{Results}
\label{sec:results}

\subsection{Dust continuum}
\label{sec:results_dust}

We show naturally-weighted dust continuum cutouts of our four targets in Figure \ref{fig:dustContinuum}, overplotted on \textit{JWST}/NIRCam F150W imaging obtained as part of UNCOVER and retrieved from the DAWN \textit{JWST} Archive \citep[DJA;][]{valentino2023}. No continuum emission is detected at the $>3\sigma$ level in any of our targets, down to an RMS of $\sigma = 9.6 - 18\,\mu\mathrm{Jy/beam}$ (Table \ref{tab:observations}), and we therefore adopt $3\sigma$ upper limits on their continuum flux densities. The non-detection of dust continuum emission from UHZ1 implies the $\sim2\sigma$ continuum signal reported in \citet{fujimoto2025_dualz} was spurious.

We also explore images tapered to a coarser resolution, which are more sensitive to extended dust emission. Such extended dust has been speculated to exist in particularly UV-luminous galaxies at high redshift, where it may be efficiently ejected through outflows, possibly driven by radiation pressure \citep[e.g.,][]{ferrara2023,ziparo2023,marques-chaves2025}. We note, however, that several other explanations for the lack of dust at $z>8$ have also been offered in the literature, and we discuss this in further detail in Section \ref{sec:discussion_continuum}.

In the $\gtrsim1.0''$ tapered imaging, we note the presence of a weak $\sim2\sigma$ continuum signal fully co-spatial with the $z=8.51$ galaxy UNCOVER-10646 (Figure \ref{fig:continuum_UC10646_tapered} in Appendix \ref{app:dust_continuum_UC10646}). This galaxy likely hosts an outflow (Section \ref{sec:discussion_outflow}), making the possibility of extended dust emission particularly intriguing, and we return to this in Section \ref{sec:discussion_outflow_afm}. However, deeper data are needed to verify the possible dust continuum signal. 

Given the lack of a robust continuum signal in the individual targets, we also perform a simple inverse-variance weighted stack using the images $uv$-tapered to a common $\sim0.8''$ resolution. The stack attains a depth of $1\sigma =8.6\,\mu\mathrm{Jy/beam}$, but does not reveal any central continuum signal at the $>2\sigma$ level.

\subsection{\oiiil{} emission}
\label{sec:results_oiii}

\begin{figure*}
    \centering
    \includegraphics[width=0.95\linewidth]{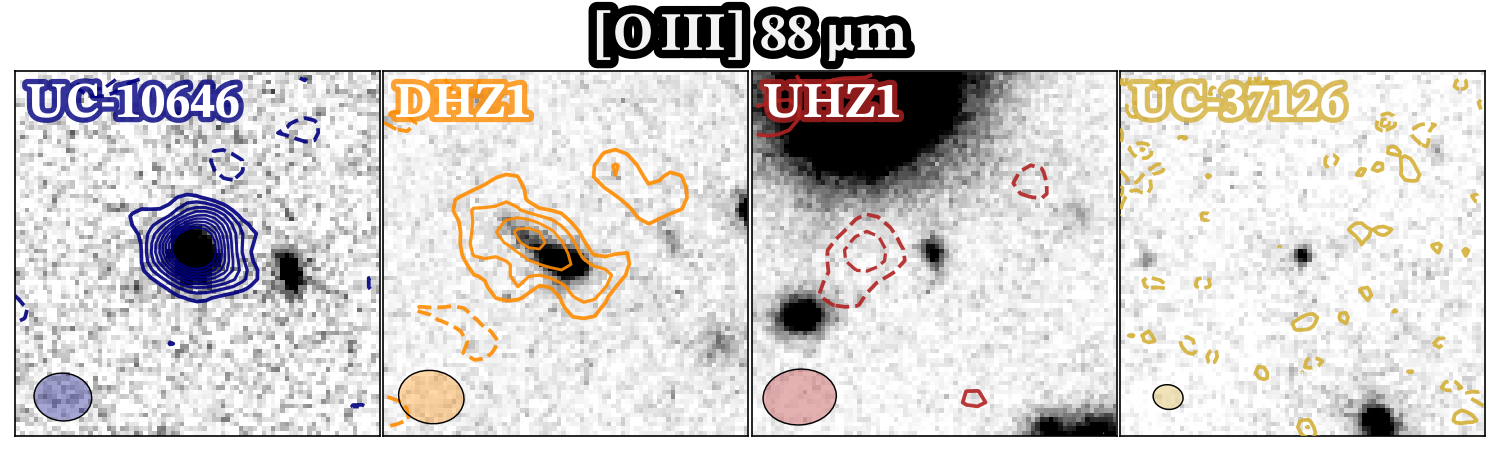}
    \caption{ALMA \oiii{} cutouts ($4''\times4''$) of our four targets, overplotted on \textit{JWST}/NIRCam F444W imaging. For UNCOVER-10646 and DHZ1, where \oiii{} emission is robustly detected, the moment-0 map spans all channels within $1.2\times$ the line FWHM. No line emission is detected in UHZ1 and UNCOVER-37126, and the moment-0 map is collapsed across $200\,\mathrm{km/s}$ centered on the fiducial redshifts of $z=10.073$ and $z=10.255$, respectively. Contours are drawn at $\pm2,3,4,\ldots, 10\sigma$ intervals, with negative contours being dashed and $\sigma$ being the RMS in the moment-0 map.}
    \label{fig:momentZero}
\end{figure*}

\begin{figure*}
    \centering
    \includegraphics[width=0.49\linewidth]{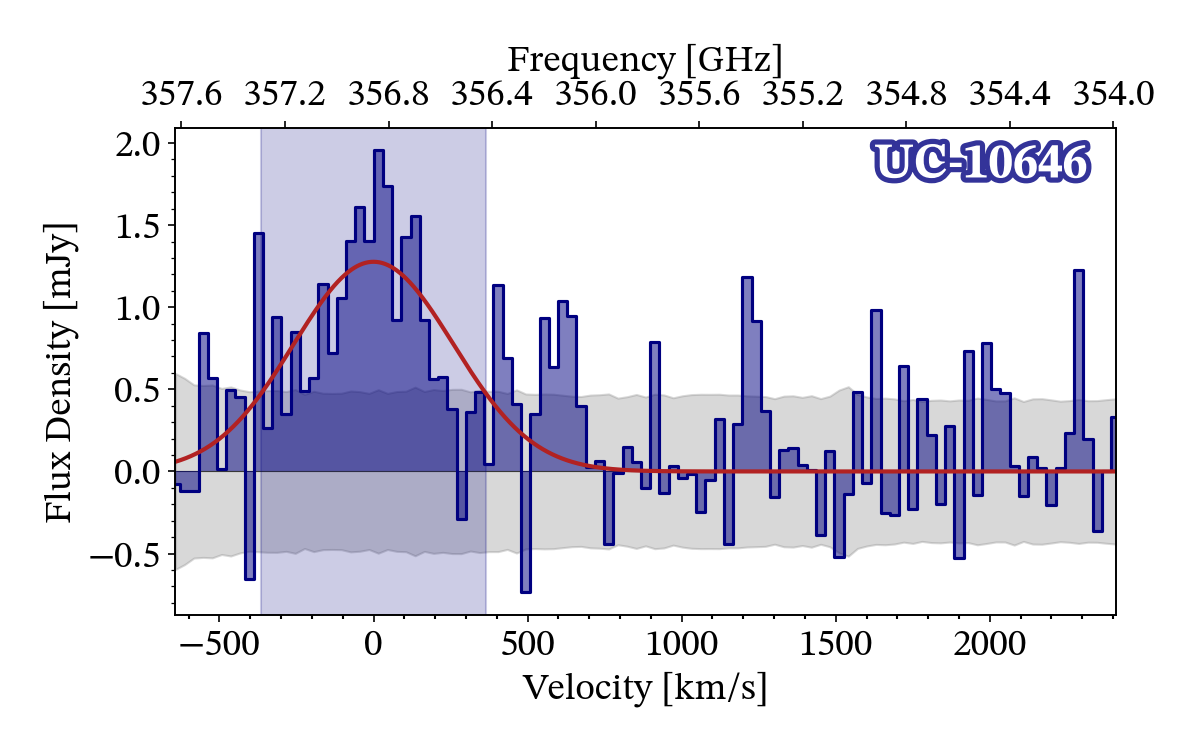}
    \includegraphics[width=0.49\linewidth]{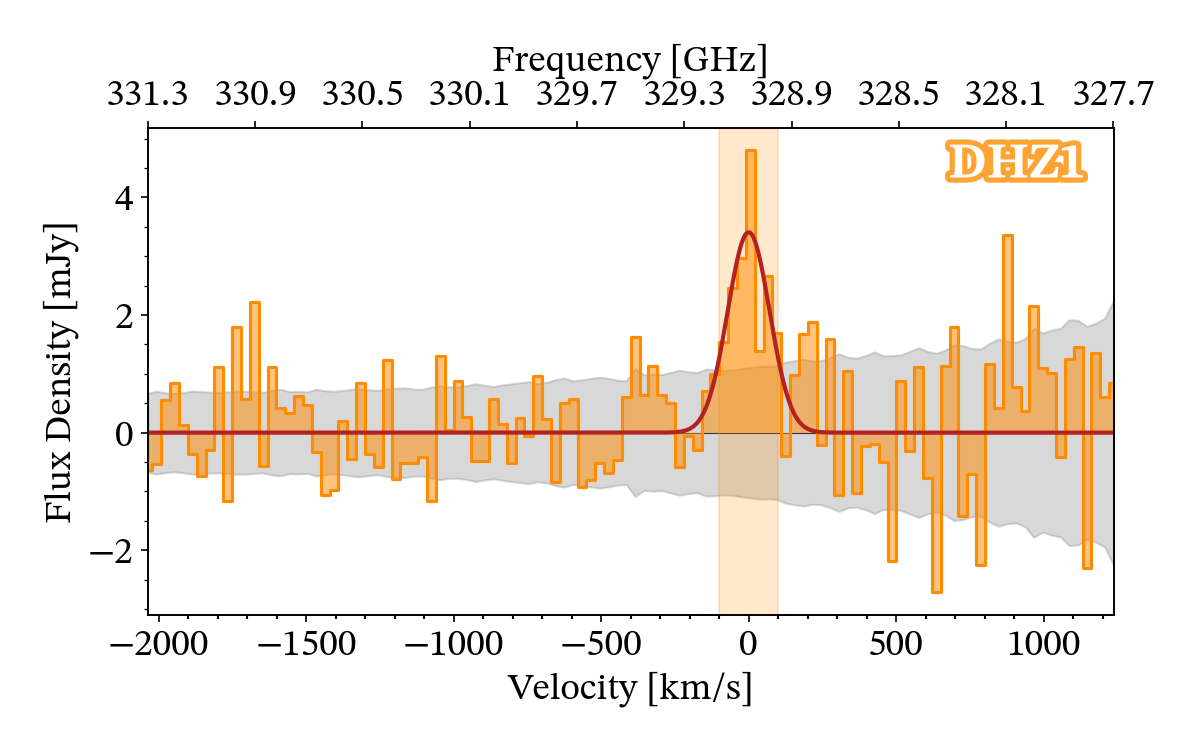}
    \caption{Extracted 1D spectra showing the \oiii{} detections for UNCOVER-10646 at $z=8.51$ (\textit{left}) and DHZ1 at $z=9.31$ (\textit{right}). A Gaussian fit to the spectra is shown through the red lines. The vertical shading shows the channels across which the moment-0 maps in Figure \ref{fig:momentZero} are collapsed, while the gray shading represents the noise per channel. The best-fit parameters are presented in Table \ref{tab:fluxes}. Note that the spectrum of UNCOVER-10646 is better fit by a double-Gaussian, as detailed in Section \ref{sec:discussion_outflow}.}
    \label{fig:spectra1D}
\end{figure*}

\begin{table*}
    \def\arraystretch{1.4}
    %\addtolength{\tabcolsep}{-0.5pt} % slightly reduce column spacing to make the table fit
    \centering
    \caption{Dust continuum, \oiiil{}, and related properties of our targets}
    \label{tab:fluxes}
    \begin{threeparttable}
    \begin{tabular}{lccccccccccc}
    
    \hline\hline
    ID & $S_\mathrm{cont}$ & $\log(M_\mathrm{dust})$ & $y_\mathrm{SN}$ & $\log(L_\mathrm{IR})$ & $\mathrm{SFR}_\mathrm{IR}$ & $f_\mathrm{obs}$ & $S_\text{\oiii{}}$ & $\mathrm{FWHM}_\text{\oiii{}}$ & $\mu L_\text{\oiii{}}$ & $L_\text{\oiii{}}$ \\
    \hline 
     & $[\mu\mathrm{Jy}]$ & $[\log(M_\odot)]$ & $[M_\odot]$ & $[\log(L_\odot)]$ & $[M_\odot\,\mathrm{yr}^{-1}]$ & & $[\mathrm{mJy\,km/s}]$ & $[\mathrm{km/s}]$ & $[10^8\,L_\odot]$ & $[10^8\,L_\odot]$\\
    \hline
    
UNCOVER-10646 (1 Gauss) & $<43.3$ & $<6.22$ & $<0.33$ & $<10.85$ & $<8.4$ & $<0.44$ & $824_{-145}^{+159}$ & $611 \pm 85$ & $22.8_{-4.0}^{+4.4}$ & $16.2_{-2.9}^{+3.1}$ \\

UNCOVER-10646 (2 Gauss) & & & & & & & $752_{-81}^{+93}$ & & $20.8_{-2.2}^{+2.6}$ & $14.9_{-1.6}^{+1.8}$ \\
\hspace*{2.5cm} Narrow & & & & & & & $251_{-57}^{+56}$ & $231_{-43}^{+46}$ & $7.0_{-1.6}^{+1.6}$ & $5.0_{-1.1}^{+1.1}$ \\
\hspace*{2.5cm} Broad & & & & & & & $507_{-87}^{+84}$ & $1366_{-329}^{+473}$ & $14.0_{-2.4}^{+2.3}$ & $10.0_{-1.7}^{+1.7}$ \\

DHZ1 & $<54.6$ & $<6.28$ & $<0.10$ & $<10.90$ & $<9.6$ & $<0.40$ & $596_{-173}^{+198}$ & $167 \pm 39$ & $18.7_{-5.4}^{+6.2}$ & $11.3_{-3.2}^{+3.7}$ \\
UHZ1 & $<28.8$ & $<5.74$ & $<0.33$ & $<10.35$  & $<2.7$ & $<0.70$ & $ < 69$ & $200$ & $< 2.4$ & $ < 0.6$ \\
UNCOVER-37126 & $<33.4$ & $< 6.06$ & $<0.23$ & $<10.67$ & $<5.7$ & $<0.64$ & $< 126$ & $200$ & $< 4.5$ & $< 2.0$ \\

\hline\hline 
\end{tabular}
\textit{Notes}: All quoted upper limits are $3\sigma$.  Col.\ (1): Galaxy identifier. Col.\ (2): upper limits on the rest-frame $\sim88\,\mu\mathrm{m}$ continuum flux density. Col.\ (3,4,5,6,7): upper limits on dust masses, supernova yields, IR luminosities, obscured SFRs, and obscured fractions $f_\mathrm{obs} = \mathrm{SFR}_\mathrm{IR} / \mathrm{SFR}_\mathrm{IR+UV}$, assuming a dust temperature of $50\,\mathrm{K}$. Col.\ (8,9): \oiii{} line flux and full-width at half maximum. Col.\ (10,11): lensed and intrinsic \oiii{} line luminosities. Values for UNCOVER-10646 are presented twice, both for the single-Gaussian fit shown in Figure \ref{fig:spectra1D}, and for the dual-Gaussian fit in Figure \ref{fig:singleDualGaussian}. The upper limits on the \oiii{} line of UNCOVER-37126 implicitly assume the line was indeed covered by the ALMA observations, though this remains unclear in light of its uncertain spectroscopic redshift (Section \ref{sec:observations_targets}).
\end{threeparttable}
\end{table*}

We detect \oiii{} at a high significance ($15\sigma$, measured as the peak pixel S/N in the moment-0 map) in UNCOVER-10646, and moreover detect the line at $6\sigma$ in DHZ1. \oiii{} remains undetected in our two $z>10$ targets. We show the moment-0 maps of the lines in Figure \ref{fig:momentZero} and the fitted 1D spectra of the two detections in Figure \ref{fig:spectra1D}. Aperture spectra of the two non-detected galaxies are shown in Figures \ref{fig:spectra1D_UHZ1} and \ref{fig:spectra1D_UC37126} in Appendix \ref{app:spectra1D_nondet}.

A Gaussian fit to the \oiii{} line from UNCOVER-10646 yields a spectroscopic redshift of $z=8.5080 \pm 0.0011$. The fit moreover yields a broad line width of $\mathrm{FWHM} = 611 \pm 85\,\mathrm{km/s}$. As shown in Section \ref{sec:discussion_outflow}, the line profile is better fit by the combination of two Gaussian components, with the broad component likely representing an outflow. The total line flux obtained from the dual-Gaussian fit is however consistent with that obtained from a single Gaussian component (Table \ref{tab:fluxes}), and we therefore proceed with this value. The de-lensed \oiii{} luminosity of UNCOVER-10646 equals $L_\text{\oiii{}} = (1.6 \pm 0.3) \times 10^9\,L_\odot$, making it the most luminous known \oiii{}-emitter at $z>8$ (Section \ref{sec:discussion_oiii_sfr}). 

The spectrum of DHZ1 is well-fit by a single Gaussian component, which accurately pins down its redshift as $z=9.3113 \pm 0.0006$. Compared to UNCOVER-10646, its \oiii{} line has a much narrower width of $\mathrm{FWHM} \approx 170\,\mathrm{km/s}$. The intrinsic line luminosity of the galaxy, after accounting for gravitational lensing, equals $L_\text{\oiii{}} = 1.1_{-0.3}^{+0.4} \times 10^9\,L_\odot$. 

Finally, we find that the \oiii{} line is not detected in UHZ1 and UNCOVER-37126, and we therefore quote an upper limit on the de-lensed line luminosity in Table \ref{tab:fluxes}, assuming a fiducial line width of $\mathrm{FWHM} = 200\,\mathrm{km/s}$. This suggests low \oiii{} luminosities of $L_\text{\oiii{}} < (0.6 - 2.0) \times 10^8\,L_\odot$ ($3\sigma$), making them at least an order of magnitude fainter than our two \oiii{}-detected targets. We discuss this further in Section \ref{sec:discussion_oiii_sfr}, in the context of the \oiii{}-star formation rate relation.

\subsubsection{Extended and spatially offset \oiii{} emission in DHZ1}
\label{sec:spatiallyOffsetOIII}

Intriguingly, the centroid of the \oiii{} emission of DHZ1 appears slightly offset from the main component of the galaxy, with the emission peaking in the extended tail identified by \citet{boyett2024}. Moreover, the galaxy appears spatially resolved in \oiii{}, and we therefore perform a 2D Gaussian fit to its moment-0 map via {\sc{casa}} {\sc{imfit}}. This yields a beam-deconvolved size for DHZ1 of $(1.47'' \pm 0.44'') \times (0.70'' \pm 0.27'')$, which at the redshift of DHZ1 corresponds to a size of $(6.4 \pm 1.9) \times (3.1 \pm 1.2)\,\mathrm{kpc}^2$. Its lensing-corrected (by a factor $1/\sqrt{\mu}$) circularized radius then equals $R = 1.7 \pm 0.4\,\mathrm{kpc}$. 

The \oiii{} line centroid -- obtained from the 2D Gaussian fit -- is offset by $\sim0.21''$ from the flux-weighted centroid of the NIRCam F444W emission from DHZ1, and by $\sim0.30''$ from the centroid of its main component (i.e., excluding the extended tail). Given that the NIRCam imaging is aligned to \textit{Gaia} \citep{valentino2023}, we assume the S/N of the ALMA data is the main source of astrometric offset. For a peak-pixel signal-to-noise of $\mathrm{S/N}\approx6$, the astrometric accuracy is approximately $\sim0.12''$, implying this offset is significant at the $1.8 - 2.5\sigma$ level.\footnote{See Equation 10.7 in the ALMA technical handbook: \url{https://almascience.eso.org/documents-and-tools/cycle12/alma-technical-handbook}} 

While the tail of DHZ1 is significantly fainter than its main component in NIRCam imaging, recent MIRI MRS observations detect the H$\alpha$ line in both, suggesting similar star formation rates on $\sim10\,\mathrm{Myr}$ timescales \citep{bik2026}. Given the correlation of \oiii{} with SFR, and possibly with burstiness \citep{algera2025_oiii}, it is likely that the \oiii{} line indeed arises from both components, yielding a centroid that is offset from the more massive but less bursty main component of DHZ1. An alternative explanation would be that the gas in the main component of DHZ1 is significantly denser, with the electron density exceeding the critical density of \oiii{}. However, based on an analysis of the resolved \oiiinowave{}$_{5007}$ / \oiii{} line ratio, \citet{bik2026} find no strong evidence that this is the case, and we therefore regard this as a less likely explanation.

From the spatial extent and FWHM of the \oiii{} line, we can determine an order-of-magnitude estimate for the dynamical mass of DHZ1. Following \citet[][their equation 2]{ubler2023}, this yields $M_\mathrm{dyn} = 1.3_{-0.6}^{+1.1} \times 10^{10}\,M_\odot$, where we have marginalized across a range of possible Sersic indices $n=0.5 - 4$ \citep[as in e.g.,][]{witstok2025}. This is $\sim4-15\times$ larger than the stellar mass of DHZ1, and thus suggests a possibly massive gas or dark matter reservoir within the galaxy, although we stress that higher resolution \oiii{} observations are required for more robust constraints on the dynamical mass of the system. 

Finally, we note that we do not attempt to infer a dynamical mass for UNCOVER-10646 given that it is not resolved in the current \oiii{} observations, and because its line FWHM likely does not represent only the gravitational potential of the galaxy, but also an outflow component (Section \ref{sec:discussion_outflow}).

\section{Discussion}
\label{sec:discussion}

\subsection{Dust production and obscured star formation}
\label{sec:discussion_continuum}

As discussed in Section \ref{sec:results_dust}, none of our targets are convincingly detected in dust continuum emission. We therefore adopt $3\sigma$ upper limits on their continuum flux densities, assuming any dust emission is contained within a single ALMA beam. We moreover assume a fiducial FIR spectral energy distribution (SED) to obtain upper limits on the dust masses $M_\mathrm{dust}$ and infrared luminosities $L_\mathrm{IR}$ of our targets. We adopt an optically thin modified blackbody with a dust temperature of $T_\mathrm{dust} = 50\,\mathrm{K}$, motivated by various model predictions at $z\approx9$ \citep{liang2019,sommovigo2022}. This is furthermore consistent with observations of galaxies at $z\approx6-8$ \citep{bakx2021,witstok2022,witstok2023_beta}, although the scatter in galaxy dust temperatures at this epoch appears large \citep[][Rajulal et al.\ in prep]{bakx2020,bakx2025,algera2024,algera2024b}. We moreover adopt a dust emissivity index of $\beta_\mathrm{IR} = 2.0$ and an opacity coefficient of $\kappa_0 = 10.41\,\mathrm{cm}^2\,\mathrm{g}^{-1}$ at $\nu_0=1900\,\mathrm{GHz}$ \citep[e.g.,][]{ferrara2022,sommovigo_algera2025}, and account for heating by and contrast against the Cosmic Microwave Background following \citet{dacunha2013}. 

These assumptions yield magnification-corrected dust masses for our sample of $\log(M_\mathrm{dust}/M_\odot) \lesssim 5.7 - 6.3$. Expressed in terms of their stellar mass, this yields a range of $M_\mathrm{dust}/M_\star \lesssim (1 - 5) \times 10^{-3}$, with the most stringent upper limit being that for the massive $z=9.31$ galaxy DHZ1. These dust-to-stellar mass ratios can be converted into upper limits on supernova (SN) yields following the framework of \citet{michalowski2015}. This implicitly assumes that additional sources of dust production, such as grain growth and AGB stars, are negligible at these early cosmic times, which may not necessarily be the case \citep[e.g.,][]{popping2017,vijayan2019,algera2026}. In principle, this means that our quoted upper limits on SN yields are conservative. Under the assumption of a \citet{chabrier2003} IMF, we then infer the SN yields of our targets to be $y_\mathrm{SN} \lesssim 0.10 - 0.33\,M_\odot\,\mathrm{SN}^{-1}$. We note that these are `effective yields', after any destruction by the reverse shock has occurred \citep[e.g.,][]{bianchi2007}.

The inferred upper limits are overall consistent with the wide range of supernova yields suggested in the literature, as compiled in the review by \citet{schneider2024}. This compilation suggests that SN yields could be as low as $y_\mathrm{SN} < 0.01\,M_\odot\,\mathrm{SN}^{-1}$ or possibly as high as $y_\mathrm{SN} \sim 1\,M_\odot\,\mathrm{SN}^{-1}$. We note that studies of UV-luminous galaxies at slightly lower redshifts ($z\approx7$) typically infer higher yields of $y_\mathrm{SN} \sim 0.1 - 1\,M_\odot\,\mathrm{SN}^{-1}$ \citep[e.g.,][]{sommovigo2022,witstok2023_beta,algera2024,algera2026}, at least when assuming all observed dust is produced solely in SNe.

One possible interpretation for the apparent lack of dust in our targets compared to UV-luminous galaxies at $z\approx7$ is that supernovae produce relatively limited quantities of dust at early times, with grain growth in the ISM producing the bulk of dust once the galaxy is sufficiently enriched with metals \citep[e.g.,][]{asano2013,popping2017,vijayan2019,triani2020,algera2024,algera2026,choban2024,narayanan2024,burgarella2025,heintz2025_zD1,bakx2025_pixiedust}. However, the low dust-to-stellar mass ratios of high-redshift galaxies can also be the result of other mechanisms. For instance, it has been suggested that dust may be efficiently ejected from galaxies through radiation-driven outflows \citep[e.g.,][]{ferrara2023,ziparo2023,ferrara2025}. In this case, observationally inferred SN yields should be regarded as lower limits, given that some of the produced dust is no longer within the galaxy, and therefore quickly cools down rendering it undetectable against the CMB. 

An alternative scenario is that dust in high-redshift galaxies ($z\gtrsim8$) has different attenuation and emission properties compared to what is seen locally and at lower redshifts. Several works have indeed suggested that grain size distributions are more top-heavy in early galaxies, as expected from dust produced by supernovae \citep[e.g.,][]{markov2025,mckinney2025,narayanan2025,shivaei2025,zhao2025}. Such large dust grains are less efficient at absorbing UV radiation, and consequently result in galaxies that are brighter in the UV yet fainter in the far-infrared for a given dust mass.

With the current data, we cannot distinguish between these possibilities. However, we address the scenario where dust is ejected from galaxies through outflows in more detail in Section \ref{sec:discussion_outflow}, in light of the detection of an ionized outflow in UNCOVER-10646. \\

Our fiducial assumptions on the dust SEDs of our targets also yield constraints on their $8-1000\,\mu\mathrm{m}$ infrared luminosities, which we find to span the range $\log(L_\mathrm{IR}/L_\odot) < 10.35 - 10.90$ (Table \ref{tab:fluxes}). Unless their dust temperatures are well above the assumed $T_\mathrm{dust} = 50\,\mathrm{K}$, this implies our targets all have $L_\mathrm{IR} < 10^{11}\,L_\odot$, making them sub-LIRGs. We convert their infrared luminosities into obscured star formation rates via $\mathrm{SFR}_\mathrm{IR}/[M_\odot\,\mathrm{yr}^{-1}] = 1.2\times10^{-10} (L_\mathrm{IR}/L_\odot)$, following \citet{inami2022}. This yields a range of upper limits on their obscured SFRs of $\mathrm{SFR}_\mathrm{IR} < 2.7 - 9.6\,M_\odot\,\mathrm{yr}^{-1}$.

We next compute upper limits on the obscured fractions of our targets as $f_\mathrm{obs} = \mathrm{SFR}_\mathrm{IR} / (\mathrm{SFR}_\mathrm{UV} + \mathrm{SFR}_\mathrm{IR})$. The UV-based SFRs are determined from the absolute UV magnitudes in Table \ref{tab:properties} and adopt a conversion factor of $\mathrm{SFR}_\mathrm{UV}\,[M_\odot\,\mathrm{yr}^{-1}] = 7.1 \times 10^{-29} \times L_\nu\,[\mathrm{erg}\,\mathrm{s}^{-1}\,\mathrm{Hz}^{-1}]$ as in \citet{bouwens2022}. This assumes that the UV emission from our targets is fully due to star formation, and implies unobscured SFRs in the range $\mathrm{SFR}_\mathrm{UV} = 1.16 - 14.3\,M_\odot\,\mathrm{yr}^{-1}$. Combined with the upper limits on their infrared-based SFRs, this yields obscured fractions of $f_\mathrm{obs} < 0.40 - 0.70$. For comparison, UV-luminous galaxies at $z\approx7$ from the REBELS survey show typical obscured fractions of $f_\mathrm{obs} \approx 0.3 - 0.6$ \citep[][see also \citealt{inami2022,bowler2024,fisher2025b}]{algera2023}. The current upper limits on $f_\mathrm{obs}$ for our $z > 8$ targets are therefore not stringent enough to rule out similarly high obscured fractions. However, given that at low redshifts $f_\mathrm{obs}$ correlates strongly with stellar mass \citep[e.g.,][]{whitaker2017}, and that the stellar masses of our $z > 8$ targets are up to $\sim1\,\mathrm{dex}$ lower than those of $z\sim7$ REBELS galaxies \citep[which have $\log(M_\star/M_\odot)\approx9 - 10$; e.g.,][]{topping2022,rowland2025}, we would expect them to be less obscured on average.

We note, however, that the upper limit on the total UV+IR star formation rate of UNCOVER-10646, $\mathrm{SFR}_\mathrm{UV+IR} \lesssim 19\,M_\odot\,\mathrm{yr}^{-1}$, is lower than the value obtained from SED fitting of $\mathrm{SFR}_\mathrm{SED} = 45.4 \pm 1.7\,M_\odot\,\mathrm{yr}^{-1}$ by Weaver et al.\ (in preparation; Table \ref{tab:properties}). This could suggest its dust temperature to be larger than the currently assumed $T_\mathrm{dust}=50\,\mathrm{K}$, or that the level of dust attenuation is overestimated based on its NIRSpec prism spectrum. A more detailed analysis of the UV and IR properties of $z>8$ galaxies with deep ALMA observations, combining the four sources analyzed in this study with the sample from \citet{bakx2025_pixiedust} based on archival ALMA data, will be the subject of a future work. Based on the current data, however, we conclude that the lack of a dust continuum detection in our four $z=8.5 - 10.3$ targets suggests they contain limited amounts of dust and host little obscured star formation.

\subsubsection{A lack of dust continuum in UHZ1}
\label{sec:results_dust_UHZ1}

The lack of a dust continuum detection in UHZ1, the $z=10.07$ X-ray AGN \citep{goulding2023,bogdan2024}, is somewhat surprising. The \textit{Chandra} X-ray detection presented in \citet{bogdan2024} suggests the AGN is heavily obscured, with an estimated $N_H = 2 \times 10^{24}\,\mathrm{cm}^{-2}$. While the fraction of heavily obscured AGN is indeed found to increase towards high redshift \citep[e.g.,][]{ueda2014,aird2015,vito2018,vijarnwannaluk2022}, this could (in part) be due to an increase in galaxy-wide obscuration, rather than just enhanced obscuration close to the central black hole \citep[e.g.,][]{trebitsch2019,ni2020,gilli2022}. 

Adopting the $A_V = 0.08$ measured by \citet{goulding2023}, this suggests a particularly low $\log(A_V / N_H) \approx -25.4\,\log(\mathrm{mag}\,\mathrm{cm}^{2})$ when compared to local galaxies \citep[$\log(A_V/N_H) \sim -22$; ][]{konstantopoulou2024} as well as those at $z\sim2 - 14$ \citep[$-24 \lesssim \log(A_V/N_H) \lesssim -21$; ][]{heintz2025_zD1,rowland2025_dla}. Moreover, \citet{goulding2023} manage to simultaneously fit the Lyman break and emission lines in UHZ1 without the need for a DLA component, implying no large HI column density in its ISM and CGM. Based on the analysis in \citet{rowland2025_dla} for $z\sim7$ galaxies, a lack of such a DLA signature in NIRSpec prism data implies $\log(N_\mathrm{H}/\mathrm{cm}^2) \lesssim 22 - 23$. We note that, at higher redshifts, a fixed HI column density yields a larger offset between the redshift inferred from the Lyman break and the true redshift as obtained from emission lines, implying this upper limit is conservative. Both the lack of an ALMA dust continuum detection and the low attenuation inferred from its \textit{JWST}/NIRSpec prism data thus suggest a lack of dust in the wider ISM of UHZ1, implying the obscuration must occur close to its nuclear region. However, we note that recent works have placed the AGN nature of UHZ1 into question \citep{alvarez-marquez2026,zou2026}, suggesting it is a metal-poor star-forming galaxy rather than a heavily obscured X-ray AGN; this would make the lack of observed dust emission less surprising.

\subsection{\oiii{}-SFR relation}
\label{sec:discussion_oiii_sfr}

\begin{figure*}
    \centering
    \includegraphics[width=0.49\linewidth]{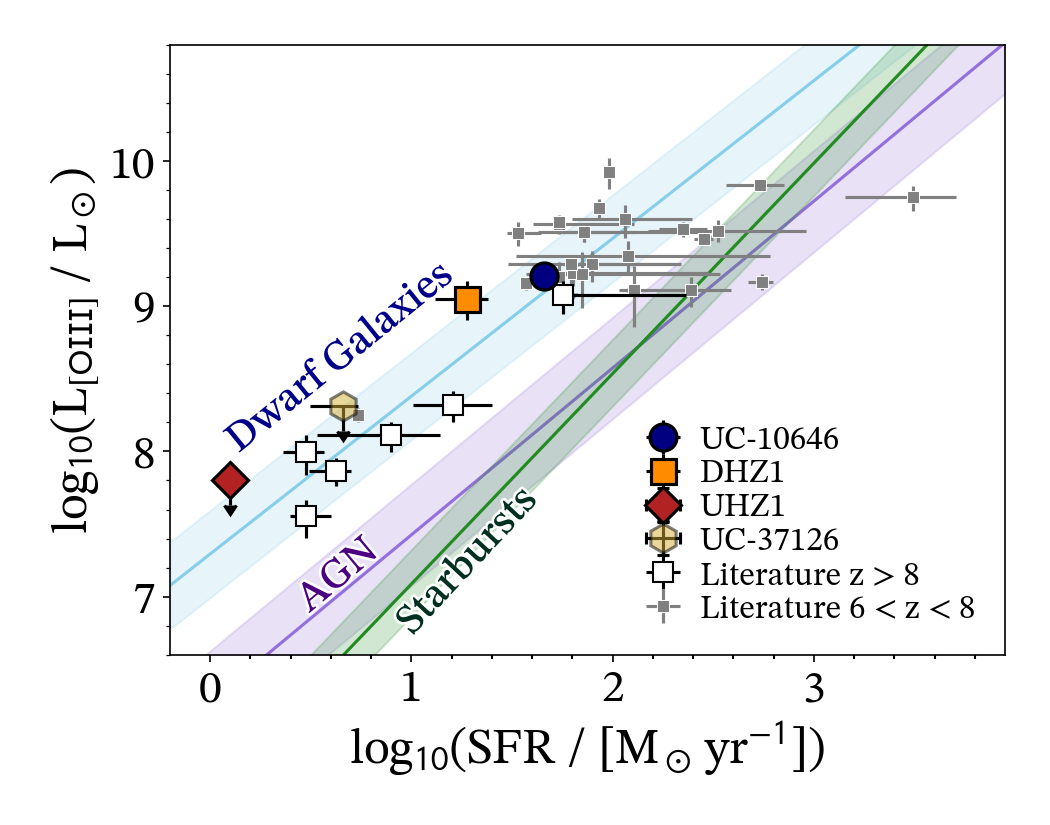}
    \includegraphics[width=0.49\linewidth]{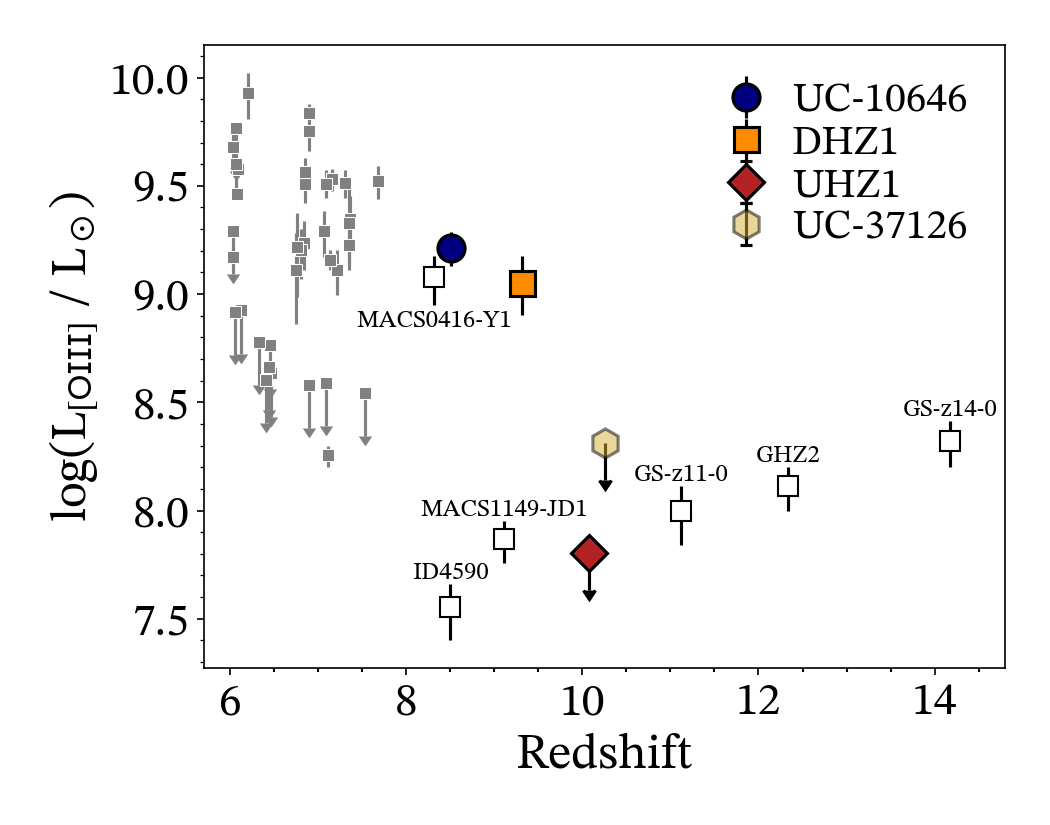}
    \caption{\textit{Left:} The \oiii{}-SFR relation of $z>6$ galaxies. Our four $z>8.5$ targets are overplotted as the colored symbols, while $z>8$ ($6 \lesssim z < 8$) galaxies compiled from the literature by \citet{algera2025_oiii} are shown as white (grey) squares. We show UNCOVER-37126 as a semi-transparent symbol, given that its \oiii{} line is possibly missed by our ALMA observations as a result of its uncertain, Lyman-break-based redshift. The local relations for starbursts (\textit{green}), dwarf galaxies (\textit{blue}) and AGN (\textit{purple}) from \citet{delooze2014} are indicated through the shaded bands. All values are corrected for lensing, where necessary. Our sample shows \oiii{} luminosities consistent with the local dwarf galaxy relation, in agreement with previous studies of $z\gtrsim6$ galaxies with $\mathrm{SFR} \lesssim 100\,M_\odot\,\mathrm{yr}^{-1}$. \textit{Right:} the \oiii{} luminosity of $z>6$ galaxies plotted against redshift. We highlight the individually \oiii{}-detected galaxies at $z>8$ \citep[c.f.,][]{hashimoto2018,tamura2019,fujimoto2024_s4590,zavala2024_alma,carniani2024_oiii,schouws2024,witstok2025}. Our line-detected galaxies, UNCOVER-10646 and DHZ1, are the two most \oiii{}-luminous sources known at $z>8$ and $z>9$, respectively.}
    \label{fig:oiiiSFR}
\end{figure*}

Locally, the \oiii{} luminosities of both dwarf and starburst galaxies are known to tightly correlate with star formation rate \citep[e.g.,][see also the recent review by \citealt{decarli_diazsantos2025}]{delooze2014,cormier2015}. This correlation moreover appears to hold for galaxies hosting an AGN \citep{delooze2014,peng2025a}. Observations of high-redshift galaxies ($z\gtrsim6$) suggest that most are consistent with the local \oiii{}-SFR relation for dwarf galaxies \citep[][]{hashimoto2018,harikane2020,zavala2024_alma,algera2025_oiii,witstok2025}, except perhaps at particularly high SFRs ($\mathrm{SFR} \gtrsim 200\,M_\odot\,\mathrm{yr}^{-1}$; \citealt{harikane2020,algera2025_oiii}). This suggests that \oiii{} is a powerful SFR tracer in the early Universe, although detections of the line at $z\gtrsim8$ remain limited.

We compile SFR measurements of our four targets from the literature (Section \ref{sec:observations_targets}; see Table \ref{tab:properties}), and show them in the context of the \oiii{}-SFR relation in the left panel of Figure \ref{fig:oiiiSFR}. For comparison, we overplot the recent compilation of $z>6$ galaxies targeted in \oiii{} from \citet{algera2025_oiii}, highlighting the subset of six $z>8$ literature sources through larger markers. We moreover compare to the canonical \oiii{}-SFR relations for local dwarfs, starburst galaxies, and AGN from \citet{delooze2014}.

Under the assumption that \oiii{} is emitted predominantly from \hii{}-regions, it is expected to trace star formation on a typical timescale of $\lesssim10\,\mathrm{Myr}$, similar to recombination lines and free-free continuum. The SFRs of our targets were obtained from SED-fitting, and trace star formation on slightly longer timescales ($\approx30-100\,\mathrm{Myr}$; Section \ref{sec:observations_targets}). Similarly, the local \citet{delooze2014} relations rely on GALEX FUV and \textit{Spitzer}/MIPS $24\,\mu\mathrm{m}$ emission to probe unobscured and obscured star formation, respectively, which trace star formation on typical timescales of $\lesssim 100\,\mathrm{Myr}$ \citep[e.g.,][]{kennicutt2012}. To further homogenize the SFR measurements of our targets and enable a more apples-to-apples comparison with the local relations, we therefore also consider UV-only and UV+IR-based SFRs in Appendix \ref{app:oiii_SFR} (Figure \ref{fig:oiiiSFR_UV_IR}). The latter constitute an upper limit on the total SFR of our targets given that no far-infrared dust emission is directly detected. Adopting these SFRs does not significantly affect our interpretation of the \oiii{}-SFR relation of our $z > 8$ sample. \\

Using the fiducial SFRs derived from SED-fitting, we find that our two \oiii{}-detected targets are consistent with the local \oiii{}-SFR relation for dwarf galaxies from \citet{delooze2014}, in agreement with other high-redshift samples. Indeed, all currently detected $z>8$ galaxies appear to fall within the $\sim0.3\,\mathrm{dex}$ scatter of this local relation \citep[c.f.,][]{hashimoto2018,tamura2019,fujimoto2024_s4590,zavala2024_alma,carniani2024_oiii,schouws2024,witstok2025} which reinforces the reliability of \oiii{} as a star formation rate tracer at early cosmic times. The upper limits for our two $z>10$ targets are similarly consistent with them being located on the local \oiii{}-SFR relation, although deeper data are needed to verify this. Moreover, we recall that the spectroscopic redshift of UNCOVER-37126 is based solely on the Lyman break, which therefore may have caused the \oiii{} line to fall outside our frequency coverage (Section \ref{sec:observations_targets}).

The fact that $z > 8$ galaxies appear to closely follow the local \oiii{}-SFR relation is noteworthy. As discussed, \oiii{} is mostly sensitive to star formation on short timescales ($\lesssim10\,\mathrm{Myr}$), while the local relations were established using SFR measurements tracing a longer typical timescale ($\lesssim100\,\mathrm{Myr}$). For a relatively constant star formation history, this would not lead to any offsets, however, high-redshift galaxies are generally believed to be more `bursty' than their lower redshift counterparts \citep[e.g.,][]{clarke2024,endsley2025_bursty,fisher2025b,munoz2026}. As a result, short term enhancements in their SFR would be expected to lead to an increase in the \oiii{} luminosity relative to the local \oiii{}-SFR relation. Indeed, recent modeling by \citet{witstok2025} suggests that a particularly young starburst aged $\sim1\,\mathrm{Myr}$ would drive a galaxy towards higher $L_\text{\oiiinowave{}}/\mathrm{SFR}$ ratios than the \citet{delooze2014} relation for dwarf galaxies, although by $\sim10\,\mathrm{Myr}$ the local relation is recovered. A secondary dependence on metallicity moreover exists, as particularly low oxygen abundances ($Z \lesssim 0.1\,Z_\odot$) may suppress the \oiii{} luminosity \citep[][see also \citealt{harikane2020}]{witstok2025}, which could thus partially compensate for the increased \oiii{} emission associated with young, bursty star formation.

Empirically, \citet{algera2025_oiii} also do not find a clear correlation between $L_\text{\oiiinowave{}}/\mathrm{SFR}$ and the equivalent width of the \oiiidopt{} and H$\beta$ lines -- a proxy for bursty star formation that tightly correlates with specific star formation rate (sSFR) -- neither for local dwarfs, nor for galaxies at $z>6$. Overall, this may therefore suggest that the current sample of \oiii{}-detected $z\gtrsim8$ galaxies is not particularly bursty, or that the effects of bursty star formation on the \oiii{}-SFR relation are limited in general. \\

Finally, in the right panel of Figure \ref{fig:oiiiSFR} we show the \oiii{} luminosities of $z>6$ \oiii{}-detected galaxies as a function of redshift. Even after accounting for gravitational lensing, our two detections have line luminosities of $L_\text{\oiii{}} \approx (1.1 - 1.6)\times 10^9\,L_\odot$ making them some of the most \oiii{}-luminous galaxies currently known at $z>8$. Specifically, UNCOVER-10646 is the most luminous \oiii{}-emitter observed thus far at $z > 8$ \citep[c.f., the slightly fainter MACS0416-Y1 at $z=8.31$;][]{tamura2019}, and DHZ1 is the brightest one at $z > 9$. Our upper limit for UHZ1, on the other hand, is one of the most stringent limits to date at this epoch, aided by its gravitational lensing magnification of $\mu\sim4$. Altogether, our data further underscore the detectability of the \oiii{} line in sufficiently bright high-redshift galaxies within reasonable integration times with ALMA.

\subsubsection{Possible impact of an AGN on the \oiii{}-SFR relation}

We recall that two of our targets possibly host an AGN: UNCOVER-10646 was identified by \citet{treiber2025} as a likely Type-2 AGN based on the detection of strong UV lines (see also Weaver et al.\ in preparation), while UHZ1 is possibly associated to a \textit{Chandra} X-ray source  \citep{bogdan2024}. We therefore briefly discuss what effect -- if any -- the presence of an AGN may have on the \oiii{}-SFR relation.

AGN are often characterized by enhanced optical \oiiiopt{} emission with respect to star-forming galaxies, which is believed to mainly originate from the narrow-line region (NLR) surrounding the central supermassive black hole \citep[e.g.,][]{netzer2015}. While emitted by the same ionic species, AGN are not expected to similarly significantly boost the far-infrared \oiii{} line. This is primarily due to the density of the NLR generally being $n_e > 10^3\,\mathrm{cm}^{-1}$ \citep{netzer1990,kakkad2018,cerqueira-campos2021}, which well exceeds the critical density of the \oiii{} line of $n_\mathrm{crit} \approx 510\,\mathrm{cm}^{-3}$ and thus suppresses it through collisional de-excitation. 

Indeed, in the local Universe the \oiii{}-SFR relation of AGN and composite sources appears to match that of starburst galaxies \citep[][see also Figure \ref{fig:oiiiSFR}]{delooze2014}, both of which are less \oiii{}-luminous than dwarf galaxies for a given SFR. Similarly, high-redshift quasars do not show any evidence for elevated \oiii{} emission \citep{walter2018,hashimoto2019_quasar}, and if anything show lower $L_\text{\oiii{}}/L_\mathrm{IR}$ ratios than the UV-bright $z\approx6-8$ galaxy population. Altogether this thus suggests that \oiii{} likely mostly traces the lower density star-forming ISM even in AGN host galaxies \citep[see also][]{harikane2025,usui2025}. As a result, we do not expect the presence of an AGN to affect the interpretation of the \oiii{} detection and non-detection of UNCOVER-10646 and UHZ1, respectively. For the former, its high star formation rate has likely rendered the line detectable irrespective of its AGN nature, while the significantly lower SFR of the latter implies our upper limit remains consistent with the local \oiii{}-SFR relation for dwarf galaxies.

\subsection{An ionized outflow in UNCOVER-10646}
\label{sec:discussion_outflow}

A Gaussian fit to the \oiii{} emission from UNCOVER-10646 yields a broad line with a $\mathrm{FWHM}\approx600\,\mathrm{km/s}$ (Section \ref{sec:results_oiii}). However, inspecting the line profile in Figure \ref{fig:spectra1D} suggests a single Gaussian component may not be sufficient to describe the emission, which appears to show broad wings, as well as residual flux in excess of the fit at $+600\,\mathrm{km/s}$. Such emission at high velocities could suggest the presence of an ionized outflow, akin to those now frequently observed for high-redshift galaxies as a broad component in the rest-optical \oiiiopt{} line \citep[e.g.,][]{llerena2023,carniani2024_outflow,xu2025}. Similar broad components appear relatively rare in \oiii{}, although a possible outflow signature in the line was reported by \citet{akins2022} for the gravitationally-lensed $z=7.13$ galaxy A1689-zD1. Unlike for UNCOVER-10646, however, the emission in this galaxy extends to velocities of only $\pm200-300\,\mathrm{km/s}$, and the outflow interpretation is unclear in light of new high-resolution \oiii{} data on A1689-zD1 suggesting a clumpy and possibly merging nature \citep{knudsen2025}. The presence of a possible broad component in \oiii{} has also previously been suggested for the $z=7.21$ Lyman-break galaxy SXDF-NB1006-2 \citep{inoue2016,ren2023}, although the modest S/N of the data prevented a conclusive detection. However, follow-up high-resolution \textit{JWST} spectroscopy of SXDF-NB1006-2 has now firmly detected a broad component ($v_\mathrm{broad}\approx630\,\mathrm{km/s}$) in the \oiiiopt{} line \citep{ren2025}, confirming the galaxy indeed hosts an ionized outflow.

While outflows are thus commonly observed in \oiiiopt{}, the brief above discussion -- as well as the broader discussion on the presence or lack thereof  of \cii{} outflows in the high-redshift literature \citep[e.g.,][]{bischetti2019,novak2020,meyer2022,sawamura2025,spilker2025} -- highlights the difficulties in robustly detecting and characterizing broad emission line wings in ALMA data. We discuss these caveats in Section \ref{sec:discussion_outflow_caveats}, but first proceed by investigating the \oiii{} line profile of UNCOVER-10646 in further detail.

\subsubsection{Fitting and attempting to resolve the high-velocity emission}
\label{sec:discussion_outflow_fitting}

To investigate the nature of the putative high-velocity emission seen UNCOVER-10646, we first attempt to spatially resolve the emission through re-imaging the \oiii{} data at higher resolution using Briggs-weighted imaging \citep{briggs1995}. In doing so, we also adopt a coarser channel width of $50\,\mathrm{km/s}$, to increase the S/N per channel. Further details on the imaging are presented in Appendix \ref{app:outflow_UC10646}. We fit the Briggs-weighted moment-0 map with a 2D Gaussian using {\sc{CASA}} {\sc{imfit}} (Figure \ref{fig:UC10646_imfit}), but find that the emission is point-like, with a maximum extent of $(< 0.36'') \times (< 0.22'')$. In what follows, we will therefore conservatively assume a fiducial size of $0.3'' \pm 0.1''$ for UNCOVER-10646, noting that a smaller size implies a higher mass outflow rate (Section \ref{sec:discussion_outflow_properties}). 

\begin{figure*}
    \centering
    \includegraphics[width=0.85\linewidth]{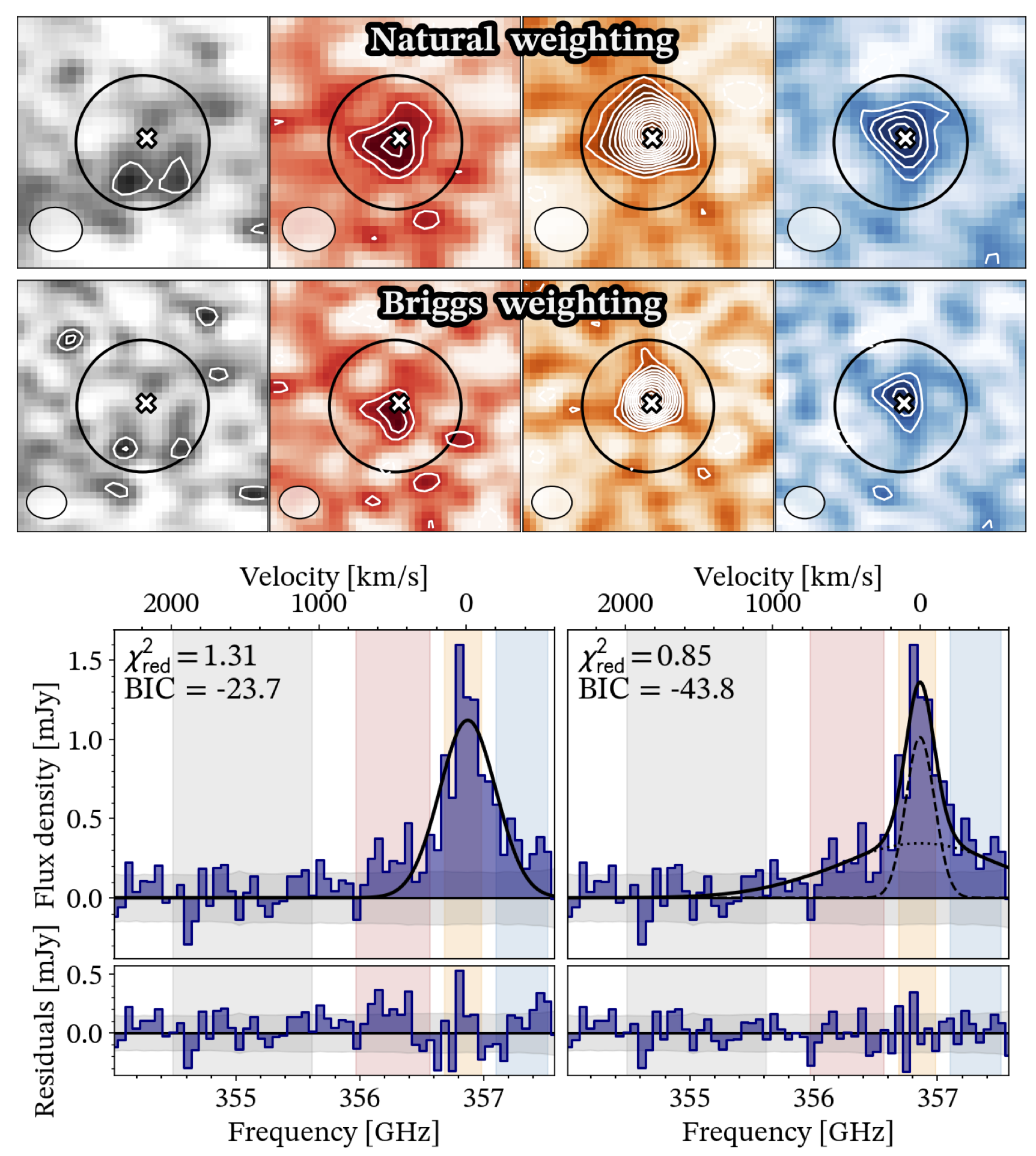}
    \caption{Naturally-weighted (\textit{top row}) and briggs-weighted (\textit{middle row}) moment-0 maps of UNCOVER-10646 extracted across the shaded regions shown in the 1D peak-pixel spectra (\textit{bottom row}). Contours are drawn at $\pm 2,3,4,\ldots\times\sigma$ intervals, and the aperture (black circle) has a radius of $0.8''$. The grey and colored maps correspond to a region of (undetected) continuum emission and the \oiii{} line, respectively. The orange shading represents the main \oiii{} component, while the blue and red shadings correspond to the blue- and redshifted emission from a possible outflow. Excess emission is seen in the red and blue components at the $>4\sigma$ level in the naturally-weighted moment-0 maps, and at $>3\sigma$ in the Briggs-weighted ones. The \textit{bottom row} shows the peak-pixel extraction (purple) from the naturally-weighted cube at the \oiii{} line center of UNCOVER-10646 in both panels (white cross in the top two panels). Single (\textit{bottom left}) and dual-Gaussian (\textit{bottom right}) fits to the spectra are shown, with the dotted and dashed curves in the right panel representing the broad and narrow Gaussian, respectively. The corresponding reduced-$\chi^2$ and BIC values of the fits are annotated in the upper left corner (note that a model with a lower BIC is preferred) and the residuals are shown below the main panels. The line profile is better described by a double-Gaussian, suggesting the presence of an ionized outflow in UNCOVER-10646.}
    \label{fig:singleDualGaussian}
\end{figure*}

Given that we do not robustly resolve the \oiii{} emission from UNCOVER-10646, we proceed by analyzing the emission in a spatially-unresolved manner. We extract a 1D spectrum from the peak-pixel of the \oiii{} emission, obtained from the naturally-weighted datacube, now also binned to $50\,\mathrm{km/s}$ channels, and show the spectrum in the bottom panels of Figure \ref{fig:singleDualGaussian}. Compared to the original aperture extraction, which adopted a $0.8''$ aperture radius, the peak-pixel spectrum more clearly highlights the presence of high-velocity positive emission at $\pm 500\,\mathrm{km/s}$. If indeed due to an outflow, this suggests the outflow is compact, and/or directed along our line of sight. A more detailed comparison between the peak-pixel and aperture spectra is shown in Figures \ref{fig:comparisonPeakApertureSpectra} and \ref{fig:singleDualGaussianApertureSpectra} in Appendix \ref{app:outflow_UC10646}. There, we also present channel maps highlighting the spatial distribution of the \oiii{} emission in $50\,\mathrm{km/s}$ bins (Figure \ref{fig:channelMaps}). \\

We first quantify the statistical significance of the candidate outflow. We perform a double-Gaussian fit to the spectrum, adopting a Markov-Chain Monte Carlo approach using the {\sc{emcee}} library in {\sc{python}} \citep{foreman-mackey2013} in order to accurately sample the parameter space and obtain robust uncertainties in light of possible parameter degeneracies. We adopt simple flat priors on all parameters, and require the width of the broad Gaussian -- representing the outflow -- to exceed that of the narrow Gaussian representing emission from the ISM of the galaxy itself.

In an initial fit, we let the central frequencies of the two lines vary, although we find that they are consistent to within $\sim1\sigma $ ($\Delta v = v_\mathrm{broad} - v_\mathrm{narrow} = 161_{-170}^{+273}\,\mathrm{km/s}$). We note that the centroid of the broad component is difficult to precisely constrain due to the \oiii{} line of UNCOVER-10646 falling close to the edge of the spectral window. In a subsequent fit, we therefore require the centroids of the narrow and broad lines to be equal. We present the measurements for the double-Gaussian fit in Table \ref{tab:fluxes}.

To assess whether the two Gaussian components provide a better description of the \oiii{} line profile of UNCOVER-10646, we re-fit its spectrum also with a single Gaussian within the same Bayesian framework, and show both fits in lower panel of Figure \ref{fig:singleDualGaussian}. We proceed by comparing the two fits using both a simple reduced-$\chi^2$ criterion ($\chi^2_\mathrm{red}$), and the Bayesian Information Criterion (BIC). Based on the former, we find the double-Gaussian fit to be statistically preferred, given its lower $\chi^2_\mathrm{red,double} = 0.85$ compared to $\chi^2_\mathrm{red,single} = 1.31$ for the single Gaussian. Moreover, we infer $\Delta \mathrm{BIC} = 20.1$ in favor of the double-Gaussian fit, suggesting `very strong evidence' that the emission is better described by two Gaussian components based on the scale by \citet[][defined as $\Delta\mathrm{BIC} > 10$]{raftery1995}.

We show moment-0 maps across the narrow and broad-line components of the \oiii{} emission in the top and middle panels of Figure \ref{fig:singleDualGaussian}. In the naturally-weighted (Briggs-weighted) maps, emission in both the red and blue component of the broad-line wings is detected at the $4 - 5\sigma$ ($3 - 4\sigma$) level. In the Briggs-weighted maps, the red component of the outflow is tentatively offset from the source center in the southern direction, although this is consistent with being due to noise at the current S/N and spatial resolution. The blue- and redshifted components are not resolved in the naturally-weighted moment-0 maps, and their centroids are consistent with the overall \oiii{} centroid of UNCOVER-10646.

\subsubsection{Outflow properties}
\label{sec:discussion_outflow_properties}

Having established that a dual-Gaussian fit accurately represents the observed \oiii{} emission profile from UNCOVER-10646, we proceed by quantifying its properties. We find that the narrow component has a FWHM of $\mathrm{FWHM}_\mathrm{narrow} = 231_{-43}^{+46}\,\mathrm{km/s}$, which is significantly narrower than we obtained from a single-Gaussian fit. The broad component, on the other hand, has a large $\mathrm{FWHM}_\mathrm{broad} = 1366_{-329}^{+473}\,\mathrm{km/s}$. The flux in the broad component, $S_\text{\oiii{},broad} = 507_{-87}^{+84}\,\mathrm{mJy\,km/s}$ is approximately twice as large as that in the narrow component of UNCOVER-10646 (see also Table \ref{tab:fluxes}). 

Assuming, then, that the broad component represents an ionized outflow, we proceed by determining the total outflow mass, closely following the approach of \citet{carniani2015} for \oiiiopt{}. We adopt the emissivity of \oiii{} -- denoted $j_\text{\oiii{}}$, which replaces the term for \oiiiopt{} in equation 5 in \citet{carniani2015} -- from {\sc{PyNeb}} \citep{luridiana2013}, assuming an electron temperature and density of $T_e = 1.5 \times 10^4\,\mathrm{K}$ and $n_e = 300\,\mathrm{cm}^{-3}$, respectively. While the electron density of UNCOVER-10646 currently remains unconstrained -- and we quantify below how adopting a different value changes our results -- the adopted electron temperature is consistent with what is inferred by Weaver et al.\ (in preparation) using the detection of the \oiiinowave{}$_{4363}$ line in \textit{JWST}/NIRSpec. The total ionized outflow mass is then calculated to be \citep[][]{carniani2015}:

\begin{equation}
\begin{split}
    \left(\frac{M_\mathrm{out}}{M_\odot}\right) & = 
    3.86 \times 10^{-12} \times C^{-1} \times \left(\frac{n_e}{1\,\mathrm{cm}^{-3}}\right)^{-1} \\
     &\times \left(\frac{j_\text{\oiii{}}}{1\,\mathrm{erg\,s}^{-1}\,\mathrm{cm}^{3}}\right)^{-1} \left(\frac{L_\text{\oiii{}}}{L_\odot}\right) \\ 
     &\times 10^{-\left[12+\log(\mathrm{O/H})\right]} \ .
\end{split}
\end{equation}

\noindent We note that this moreover assumes that all oxygen is in a double-ionized state. We define $C=\langle n_e^2 \rangle / \langle n_e \rangle^2$ as the clumping factor, from hereon taken to be $C=1$. We note, however, that if the outflow is clumpy ($C>1$), the true outflow mass will be lower. The emissivity equals $j_\text{\oiii{}} = 6.77 \times 10^{-22}\,\mathrm{erg\,s}^{-1}\,\mathrm{cm}^3$ given our fiducial $n_e$ and $T_e$. We moreover adopt the metallicity of UNCOVER-10646 from Weaver et al.\ (in preparation), who determine $12 + \log(\mathrm{O/H}) = 8.12 \pm 0.01$, and a de-lensed \oiii{} luminosity for the broad component of $L_\text{\oiii{}} = (1.0 \pm 0.2) \times 10^9\,L_\odot$ (Table \ref{tab:fluxes}). Combined, this yields a substantial ionized outflow mass of $M_\mathrm{out} = (1.44 \pm 0.25)\times10^8\,M_\odot$. Adopting the same set of assumptions for the narrow component yields the ionized mass within the ISM of UNCOVER-10646, which we find corresponds to $M_\mathrm{ISM}^\mathrm{ion} = (7.2 \pm 1.6) \times 10^7\,M_\odot$ -- a factor of two smaller, as expected based on the flux ratios in the broad and narrow lines. Had we adopted a lower (higher) electron density of $n_e = 100\,\mathrm{cm}^{-3}$ ($500\,\mathrm{cm}^{-3}$), the inferred outflow and ISM masses would be larger (smaller) by a factor of $2\times$ ($1.2\times$). Varying the electron temperature between $T_e = (1 - 2)\times10^4\,\mathrm{K}$, on the other hand, does not change the inferred masses by more than $10\,\%$ given our fiducial electron density.

We proceed by determining the mass outflow rate of UNCOVER-10646, which can roughly be approximated by $\dot{M}_\mathrm{out} \sim M_\mathrm{out} v_\mathrm{out} / R_\mathrm{out}$ where $v_\mathrm{out}$ is the velocity of the outflow and $R_\mathrm{out}$ its size. This equation is valid both for a spherical and biconical outflow, and implicitly assumes an isothermal density profile \citep{rupke2005,carniani2015,veilleux2020}. Adopting a constant density across the outflow would yield an outflow rate that is a factor of $3\times$ larger. Regardless, the precise geometry of the unresolved outflow does not strongly affect the inferred mass outflow rate.

For the velocity, we follow \citet{carniani2024_outflow} by adopting $v_\mathrm{out} = |\Delta v| + 2\sigma_\mathrm{broad} = 2 \sigma_\mathrm{broad}$, where the latter equality is true by construction since we fixed the centroids of the broad and narrow lines in our fitting approach. However, a possible velocity offset between the lines is in any case negligible when compared to the width of the broad component, and will thus not significantly affect the inferred outflow velocity. This approach yields a high $v_\mathrm{out} = 1159_{-279}^{+401}\,\mathrm{km/s}$. In the local Universe, such high-velocity ionized outflows are generally AGN-powered \citep[e.g.,][]{fiore2017}, which is consistent with the finding by Weaver et al.\ (in preparation) that UNCOVER-10646 consists of a galaxy pair with at least one, and possibly two, AGN (Section \ref{sec:observations_targets}). However, based on a comparison to theoretical works in Section \ref{sec:discussion_outflow_comparison}, we find that a dominant AGN contribution to the outflow may not be explicitly required for UNCOVER-10646.

Determining $\dot{M}_\mathrm{out}$ requires additional information on the extent of the outflow, for which we adopt our previously estimated fiducial size of $0.3'' \pm 0.1''$, which at $z=8.51$ corresponds to $R_\mathrm{out} \approx 1.4 \pm 0.3\,\mathrm{kpc}$. From these values, we determine a high mass outflow rate of $\dot{M}_\mathrm{out} = 128_{-46}^{+80}\,M_\odot\,\mathrm{yr}^{-1}$. We note that this should be taken as a lower limit if the outflow is significantly more compact than the ALMA beam. Expressing the ionized outflow rate in terms of the total star formation rate of UNCOVER-10646 implies a high mass loading factor $\eta = \dot{M}_\mathrm{out} / \mathrm{SFR} = 2.9_{-1.0}^{+1.8}$. If only one of the two merging galaxies is driving the outflow, this too should be considered a lower limit given that this would overestimate the relevant SFR in calculating the mass loading factor. While we will proceed with $\eta \sim 3$ as our fiducial mass loading factor, the outflow rate could thus plausibly be significantly larger. Future, high-resolution ALMA follow-up is necessary to investigate this in further detail by resolving the outflow, pinpointing its host, and measuring its spatial extent.

Before placing the observed outflow in UNCOVER-10646 in the context of the literature in Section \ref{sec:discussion_outflow_comparison}, we perform a rough estimate of whether the outflowing material is expected to escape the potential well of the galaxy and thereby ultimately enrich the IGM. We follow the approach from \citet[][their equations $4 - 6$]{cooper2025}, which in turn adopts the definition of escape velocity from \citet{heckman2000}. This prescription requires an estimate of the halo mass of UNCOVER-10646, for which we assume $M_h \sim 10^{11}\,M_\odot$ based on its stellar mass (Table \ref{tab:properties}) and the stellar-to-halo mass relation \citep{behroozi2019,shuntov2025_stellar2halo}. We further require an estimate of the effective radius of UNCOVER-10646, for which we adopt $r\sim100\,\mathrm{pc}$ based on the NIRCam imaging presented in Weaver et al.\ (in preparation). We subsequently infer an escape velocity of $v_\mathrm{esc} \sim 600\,\mathrm{km/s}$, which is consistent with the escape velocities inferred for similarly massive galaxies in the literature \citep{carniani2024_outflow,cooper2025,xu2025}, and is well below the measured outflow velocity of $v_\mathrm{out} \approx 1160\,\mathrm{km/s}$ for UNCOVER-10646. This suggests that the outflow is likely to escape the galaxy and enrich the IGM, although we stress the approximate nature of this estimate.

\begin{figure}
    \centering
    \includegraphics[width=0.52\textwidth]{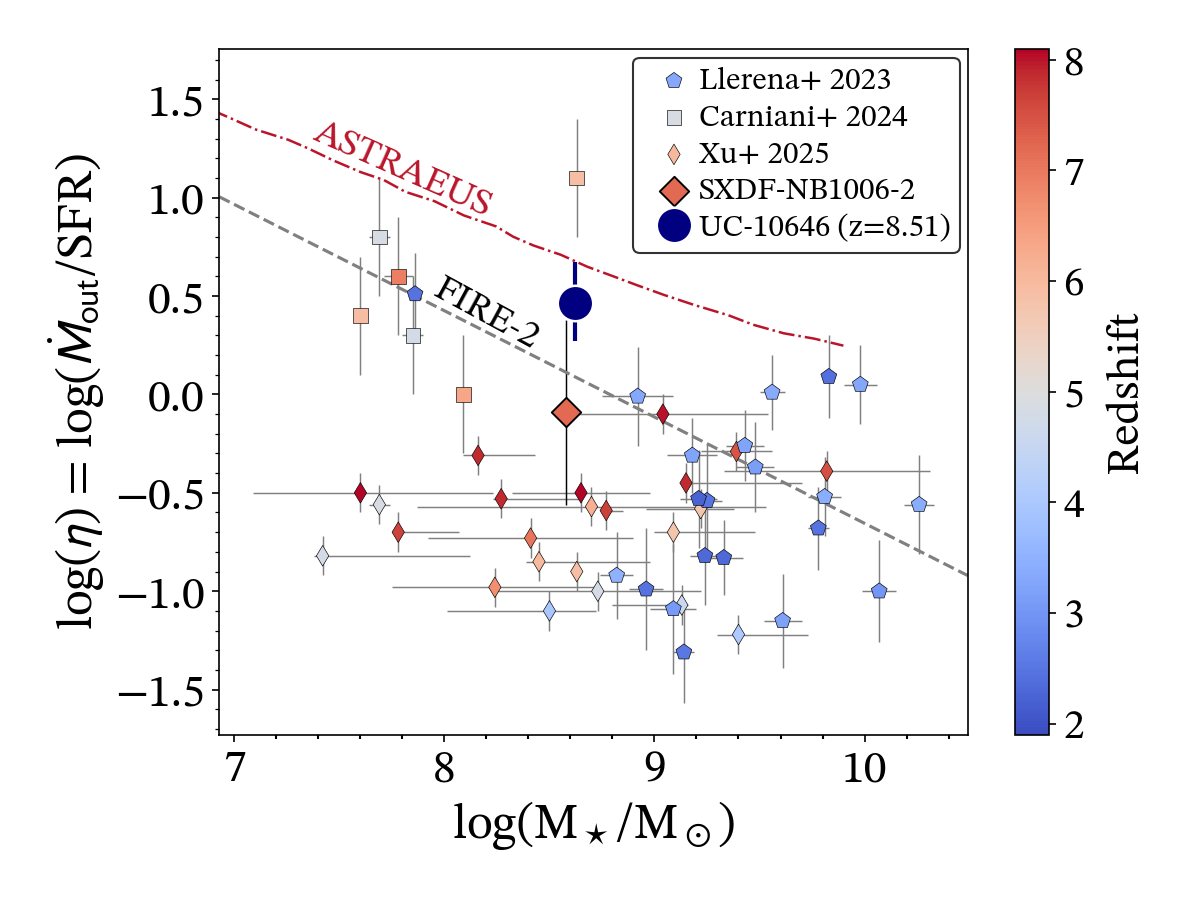}
    \caption{The mass loading factor $\eta = \dot{M}_\mathrm{out}/\mathrm{SFR}$ as a function of stellar mass. UNCOVER-10646, whose mass loading factor is derived from the \oiiil{} line, is shown as the large blue marker, and is compared to \oiiiopt{}-derived outflow rates at $z\approx2-8$ \citep{llerena2023,carniani2024_outflow,xu2025}. We also explicitly compare to the ALMA-detected $z=7.21$ galaxy SXDF-NB1006-2, for which \citet{ren2025} recently identified an \oiiiopt{}-based outflow. The predictions for ionized and total outflow rates from, respectively, FIRE-2 \citep[][independent of redshift]{pandya2021} and ASTRAEUS \citep[][at $z=8$]{ucci2023} are indicated through the dashed gray and dash-dotted red lines. UNCOVER-10646 shows a particularly powerful ionized outflow when compared to galaxies of similar stellar mass, but is in broad agreement with predictions from theoretical models.}
    \label{fig:outflowRates}
\end{figure}

\subsubsection{The outflow of UNCOVER-10646 in context}
\label{sec:discussion_outflow_comparison}

We compare the mass loading factor of UNCOVER-10646 to galaxy outflow rates inferred from the \oiiiopt{} line in Figure \ref{fig:outflowRates}. We opt to focus on outflows identified from the same ionic species for overall consistency in deriving the outflow properties between our work and the literature. Moreover, our comparison studies adopt broadly similar approaches in determining mass outflow rates as outlined for UNCOVER-10646 in Section \ref{sec:discussion_outflow_properties}, which is necessary for a reliable comparison.

We first compare to the work by \citet{llerena2023}, who study $35$ star-forming galaxies at $z\sim3$ with stellar masses of $M_\star \approx 10^{9-10}\,M_\odot$ using ground-based spectroscopy. Approximately two-thirds of their galaxy sample is found to host an ionized outflow with a typical mass-loading factor of $\eta \approx 0.5$, which is well below the value inferred for UNCOVER-10646. However, a few galaxies in the \citet{llerena2023} sample with high SFR surface densities do reach mass loading factors $\eta > 1$, similar to what is observed for our target.

We moreover compare to the recent \textit{JWST}-based studies of \oiiiopt{} outflows in $z\approx 3 - 9$ galaxies by \citet{carniani2024_outflow} and \citet{xu2025}. These works find apparently contrasting results -- on the one hand, \citet{carniani2024_outflow} argue on the basis of high-resolution NIRSpec observations that powerful ionized outflows, with mass-loading factors $\eta > 1$, are common in low-mass galaxies. On the other hand, \citet{xu2025} find typically lower values ($\eta \approx 0.1 - 1$) based on medium- and high-resolution NIRSpec data, as well as medium-resolution NIRCam/WFSS grism observations, which are more in line with mass-loading factors observed in galaxies at lower redshifts. We note, however, that while the stellar mass ranges of both studies partially overlap, the \citet{carniani2024_outflow} sample is slightly less massive, with a typical $M_\star \sim 10^8\,M_\odot$. The mass of UNCOVER-10646 of $\log(M_\star/M_\odot) \approx 8.6$ (Table \ref{tab:properties}) is more in line with the average stellar mass of the \citet{xu2025} sample. However, its high mass loading factor of $\eta \sim 3$ is $\sim1\,\mathrm{dex}$ larger than what is observed by \citet{xu2025}, and is instead more consistent with the powerful outflows inferred by \citet{carniani2024_outflow}. This suggests UNCOVER-10646 may host a particularly powerful outflow for its mass.

We also compare UNCOVER-10646 to the ALMA-detected galaxy SXDF-NB1006-2 at $z=7.21$, which was recently found to host an outflow based on a broad \oiiiopt{} line \citep{ren2025}. While only a tentative broad line was seen in its \oiii{} emission, this is not unexpected given the limited signal-to-noise of the ALMA observations \citep{inoue2016,ren2023}. We adopt the stellar mass of SXDF-NB1006-2 measured by \citet{ren2025} using a non-parametric star formation history, and adopt the mass-loading factor obtained from its SFR averaged across the last $10\,\mathrm{Myr}$. Both SXDF-NB1006-2 and UNCOVER-10646 have a similar stellar mass, and their mass-loading factors are consistent within the large (systematic) uncertainties.

Finally, we compare to the theoretical prediction on ionized outflow rates of galaxies from the FIRE-2 simulation suite \citep{pandya2021} and the ASTRAEUS semi-analytical models \citep{ucci2023}. For FIRE-2, we adopt the prediction for the `warm phase' of $T = 10^{3-5}\,\mathrm{K}$, which is expected to be traced by \oiiinowave{}. Focusing on galaxies at $z=0-4$, \citet{pandya2021} suggest that, at $M_\star \lesssim 10^{10}\,M_\odot$, most of the outflowing mass is in this warm phase. However, we caution that in simulations the bulk of the mass is generally `warm' \citep[e.g.,][]{fielding2017,hu2019,steinwandel2024}, and as such this does not necessarily imply that any colder outflow component in UNCOVER-10646 is negligible. For the semi-analytical ASTRAEUS model, which does not distinguish between different phases, we show the total outflow rates at $z=8$.

The typical mass loading factors -- both ionized and across all phases -- decrease with increasing stellar mass in both FIRE-2 and ASTRAEUS. Given the large uncertainties on the mass loading factor of UNCOVER-10646, we conclude that the inferred value of $\eta \sim 3$ is in broad agreement with both of the theoretical predictions, although could possible exceed them if the current outflow rate is being underestimated. We moreover note that neither of the theoretical works explicitly include AGN, such that the broad overall agreement with these predictions suggests that the outflow in UNCOVER-10646 could plausibly be star-formation-driven, without the need for significant energy input from a central black hole.

\subsubsection{A powerful, super-Eddington outflow?}
\label{sec:discussion_outflow_afm}

One of the most puzzling findings  from the \textit{JWST} is the high number density of UV-luminous galaxies in the early Universe \citep[e.g.,][]{bouwens2023,harikane2023,chemerynska2024_uvlf}. Among the various explanations that have been offered for this observation \citep[e.g.,][]{dekel2023,pallottini_ferrara2023,somerville2025} is the possibility that early galaxies are essentially dust-free due to radiation-driven outflows clearing the dust on very rapid timescales \citep{ferrara2023,ziparo2023,ferrara2025,ferrara2025_gsz14,nakazato2025_afm}. In this `attenuation-free model', such outflows are launched when the specific SFR (sSFR) of a galaxy exceeds a threshold of $\mathrm{sSFR} = 25\,\mathrm{Gyr}^{-1}$ \citep{ferrara2024}, known as the `super-Eddington condition'. Adopting its star formation rate and stellar mass from Table \ref{tab:properties}, we infer $\mathrm{sSFR} = 109_{-13}^{+14}\,\mathrm{Gyr}^{-1}$ for UNCOVER-10646, i.e., well in excess of the super-Eddington threshold.

It is therefore possible that the putative outflow is a result of the powerful starburst in UNCOVER-10646, which itself may have been triggered by its ongoing merger. The AGN in the galaxy may further be injecting energy, resulting in the high observed outflow velocity and mass-loading factor. It is then particularly tempting to speculate that the faint, extended continuum signal seen in tapered imaging of UNCOVER-10646 (Figure \ref{fig:continuum_UC10646_tapered} in Appendix \ref{app:dust_continuum_UC10646}) is the result of the outflow clearing its dust, as predicted by the attenuation-free model.

On the other hand, we caution that UNCOVER-10646 appears less blue than the $z\gtrsim10$ `Blue Monsters' for which the attenuation-free model was originally devised. Indeed, the UV continuum slope of the entire system of $\beta_\mathrm{UV} \approx -2.08$ suggests a non-negligible level of dust attenuation (Weaver et al.\ in preparation), although one of the two merging galaxies is found to be bluer ($\beta_\mathrm{UV} \approx -2.43$). At this stage, however, it cannot be established whether the outflow is indeed ejecting newly-formed dust from UNCOVER-10646, and higher-resolution \oiii{} observations will be invaluable to investigate this in more detail in the future.

\subsubsection{Caveats regarding the outflow interpretation}
\label{sec:discussion_outflow_caveats}

The ALMA observations of UNCOVER-10646 reveal a plausible broad component in its \oiii{} emission, consistent with the expected signature of an ionized outflow. However, while the inclusion of a broad component is indeed statistically preferred, we caveat that conclusively establishing the presence of an outflow through a broad emission line in interferometric data is notoriously difficult. In a few cases, either new observations or a re-reduction of archival data have yielded a non-detection of broad lines that previously seemed clear-cut \citep[e.g.,][]{meyer2022,sawamura2025}. In the case of \citet{sawamura2025}, for example, they demonstrate how in previous ALMA observations artificial broad-line wings in the \cii{} line arose due to a sub-optimal continuum subtraction. In our case, however, this is not an issue given the lack of bright continuum emission in UNCOVER-10646. 

We moreover verify in Appendix \ref{app:outflow_UC10646} (right panel of Figure \ref{fig:comparisonPeakApertureSpectra}) that the high-velocity emission is also present in a dirty (i.e., non-{\sc{clean}}ed) datacube. As discussed extensively in \citet{meyer2022}, as well as in several other works \citep[e.g.,][]{novak2020,czekala2021,posses2025}, {\sc{clean}}ing is generally performed only on bright emission, and down to a given flux density threshold (in our case, $2\sigma$). When creating the final {\sc{clean}} datacube, the {\sc{clean}} components and residuals from the dirty image are co-added, which can in some cases lead to mismatches in the overall flux scale -- particularly when dealing with extended emission in high-resolution data. The fact that we also see high-velocity emission in the dirty \oiii{} cube, however, suggests that it is not an artifact of our imaging procedure.

However, we note that, if real, the outflow component is necessarily compact. The signal-to-noise ratio of the broad emission is maximized in a peak-pixel extraction, and Figure \ref{fig:comparisonPeakApertureSpectra} in Appendix \ref{app:outflow_UC10646} suggests that the significance of the broad component decreases for larger aperture radii. This is expected for a fully unresolved signal, given that the noise in an aperture increases linearly with aperture radius. We quantify the significance of the broad component in our fiducial $0.8''$-radius aperture extraction in Figure \ref{fig:singleDualGaussianApertureSpectra}. In this case, a comparison of the single-and dual-Gaussian fits suggests weaker evidence for a second Gaussian component. The corresponding $\Delta\mathrm{BIC} = 2.7$ suggests only `positive evidence' for a dual-Gaussian model being preferred, per the \citet{raftery1995} scale, rather than the `very strong evidence' obtained in Section \ref{sec:discussion_outflow_fitting} for the peak-pixel spectrum ($\Delta\mathrm{BIC}\approx20$). \\

Even if the broad \oiii{} component is indeed real, that does not necessarily imply an outflow is the only possible interpretation. We recall that UNCOVER-10646 is a likely merger of two AGN based on NIRSpec prism spectroscopy (Section \ref{sec:observations_targets}). Depending on the stage of the merger, the separation between the two galaxies, and their orientation along the line of sight, this may give rise to a variety of emission line profiles. For example, a merger along the line of sight at relatively large separations is expected to yield two narrow \oiii{} emission components corresponding to the ISM of both host galaxies. However, if the radial velocity between the two merging galaxies is small -- for instance when they are nearing coalescence and/or if the merger occurs perpendicular to the line of sight -- one expects to see only a single narrow component. As we observe an approximately symmetric emission line profile on scales of $v\sim1000\,\mathrm{km/s}$ for UNCOVER-10646, this is unlikely to be the result of a single major merger system. Instead, this is consistent with both components contributing to the narrow ALMA \oiii{} emission ($\mathrm{FWHM} \approx 230\,\mathrm{km/s}$), while the broad line represents an outflow from one of the two (or possibly both). In principle, a triple merger could give rise to high-velocity emission both blue- and redwards of the narrow line, but there is no evidence for a third component in existing NIRCam imaging and NIRSpec spectroscopy. As such, we prefer the scenario where UNCOVER-10646 is a compact, late-stage merger, with the high-velocity \oiii{} emission being the result of an outflow. 

Ultimately, however, high-resolution \oiii{} imaging of UNCOVER-10646 is essential to spatially resolve the putative outflow, and isolate the individual kinematics of the two merging galaxies. At the same time, high-resolution \textit{JWST}/NIRSpec spectroscopy, beyond the $R\sim100$ offered by the current NIRSpec prism data, will be necessary to establish whether the rest-optical \oiiidopt{} lines also show a broad component.

\section{Summary and Future Outlook}
\label{sec:conclusions}

We presented the first systematic study into the \oiiil{} emission of galaxies at $z>8$, making use of new ALMA Band 7 observations towards four luminous \textit{JWST}-selected galaxies at $z=8.5 - 10.3$ benefiting from modest gravitational lensing (magnification $\mu=1.4 - 3.8$). Our main results are the following:

\begin{itemize}
    \item Two of our targets, UNCOVER-10646 at $z=8.51$ \citep[][Weaver et al.\ in preparation]{fujimoto2024_uncover} and DHZ1 at $z=9.31$ \citep{boyett2024}, are robustly detected in \oiii{} at $\sim15\sigma$ and $\sim6\sigma$, respectively. UNCOVER-10646 is the intrinsically most luminous \oiii{}-emitter currently known at $z>8$, and DHZ1 the most luminous one at $z>9$.    
    
    \item Our remaining two targets, the $z=10.07$ X-ray AGN UHZ1 \citep{goulding2023,bogdan2024} and the UV-luminous galaxy UNCOVER-37126 at $z\approx10.26$ \citep{fujimoto2024_uncover}, are not detected in \oiii{} emission. For UHZ1 in particular, this provides a deep upper limit of $L_\text{\oiii{}} < 6 \times 10^7\,L_\odot$, the deepest for a $z>10$ galaxy to date. For UNCOVER-37126, the \oiii{} line is possibly missed in the adopted ALMA tuning given an uncertain prior redshift based solely on the Lyman break.

    \item Our full sample is consistent with the local relation between \oiii{} luminosity and star formation rate for dwarf galaxies from \citet{delooze2014}. This is in agreement with other high-redshift \oiii{}-detected galaxies in the literature \citep[e.g.,][]{harikane2020,zavala2024_alma,algera2025_oiii,witstok2025}. 

    \item The \oiii{} line profile of UNCOVER-10646 is best fit by the combination of a broad + narrow Gaussian, hinting at the presence of an ionized outflow. While additional observations are needed to confirm and spatially resolve the outflow, the current data suggest a high mass outflow rate of $\dot{M}_\mathrm{out} = 128_{-46}^{+80}\,M_\odot\,\mathrm{yr}^{-1}$, which exceeds the SFR of UNCOVER-10646 by a factor $\eta = 2.9_{-1.0}^{+1.8}$. However, given the unresolved nature of the current data, this value could plausibly be larger if the outflow is particularly compact. Taken at face value, the inferred mass loading factor matches or exceeds that of recent observations and theoretical predictions of high-redshift ionized outflows \citep[e.g.,][]{pandya2021,ucci2023,carniani2024_outflow}.
    
    \item \vspace*{0.01cm} While no dust continuum emission is confidently detected in our sample, UNCOVER-10646 shows an intriguing $\sim2\sigma$ feature in imaging tapered to $>1.0''$ that could hint at an extended dust reservoir, possibly the result of its (dusty) outflow. This interpretation is consistent with predictions from the attenuation-free model \citep[e.g.,][]{ferrara2023,ferrara2024,ferrara2025}, but requires additional observations to verify.

\vspace*{0.1cm}
\end{itemize}

Our study further highlights the potential of ALMA in characterizing the earliest galaxies through the bright \oiiil{} line, building upon several pioneering studies targeting the line at $z>8$ both pre-\textit{JWST} \citep{hashimoto2018,tamura2019} and in the current \textit{JWST} era \citep{fujimoto2024_s4590,zavala2024_alma,carniani2024_oiii,schouws2024,witstok2025}. This has paved the way for the approved Cycle 12 ALMA Large Program PHOENIX, which will target \oiii{} emission in an additional 15 UV-luminous galaxies out to $z\sim15$ (2025.1.01606.L; PI Schouws), enabling statistical insights into the ISM conditions, obscured star formation, and dust build-up at Cosmic Dawn. Combined with detailed, panchromatic follow-up studies of individual high-redshift galaxies, these observations are set to significantly advance our understanding of how the earliest galaxies form and evolve.

\section*{Acknowledgments}

\noindent 

We thank the referee for their valuable and constructive comments that improved this work. We also thank Javier Álvarez-Márquez for useful discussions about the MIRI observations of UHZ1 and UNCOVER-37126. HSBA gratefully acknowledges support from Academia Sinica through grant AS-PD-1141-M01-2. MA is supported by FONDECYT grant number 1252054, and gratefully acknowledges support from ANID Basal Project FB210003 and ANID MILENIO NCN2024\_112. AKI acknowledges support from KAKENHI Grant No.\ 23H00131. NW was supported by the GACR EXPRO grant No. 21-13491X.

This paper makes use of the following ALMA data: \\
ADS/JAO.ALMA\#2024.1.00440.S, 
ADS/JAO.ALMA\#2024.1.01197.S

ALMA is a partnership of ESO (representing its member states), NSF (USA) and NINS (Japan), together with NRC (Canada), MOST and ASIAA (Taiwan), and KASI (Republic of Korea), in cooperation with the Republic of Chile. The Joint ALMA Observatory is operated by ESO, AUI/NRAO and NAOJ.

\bibliographystyle{mn2e}
\bibliography{main}

\appendix

\section{A.\ 1D spectra of UHZ1 and UNCOVER-37126}
\label{app:spectra1D_nondet}

\begin{figure}
    \centering
    \includegraphics[width=0.75\linewidth]{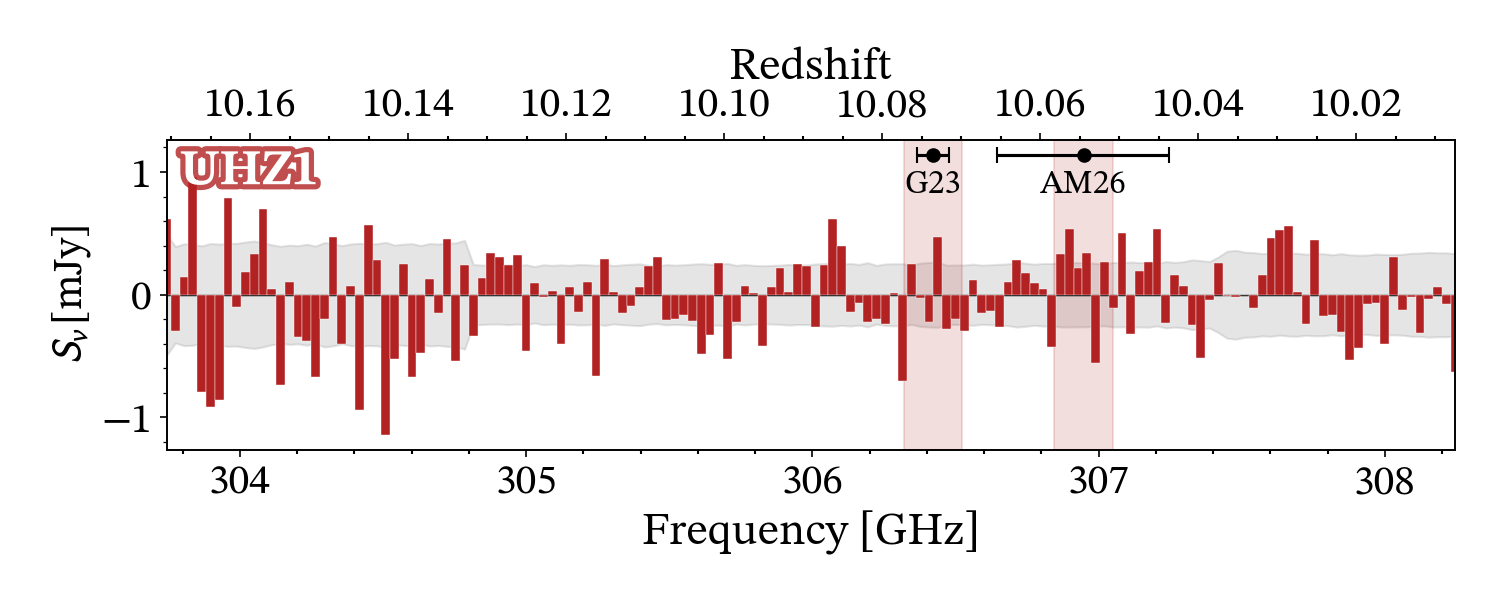}
    \includegraphics[width=0.75\linewidth]{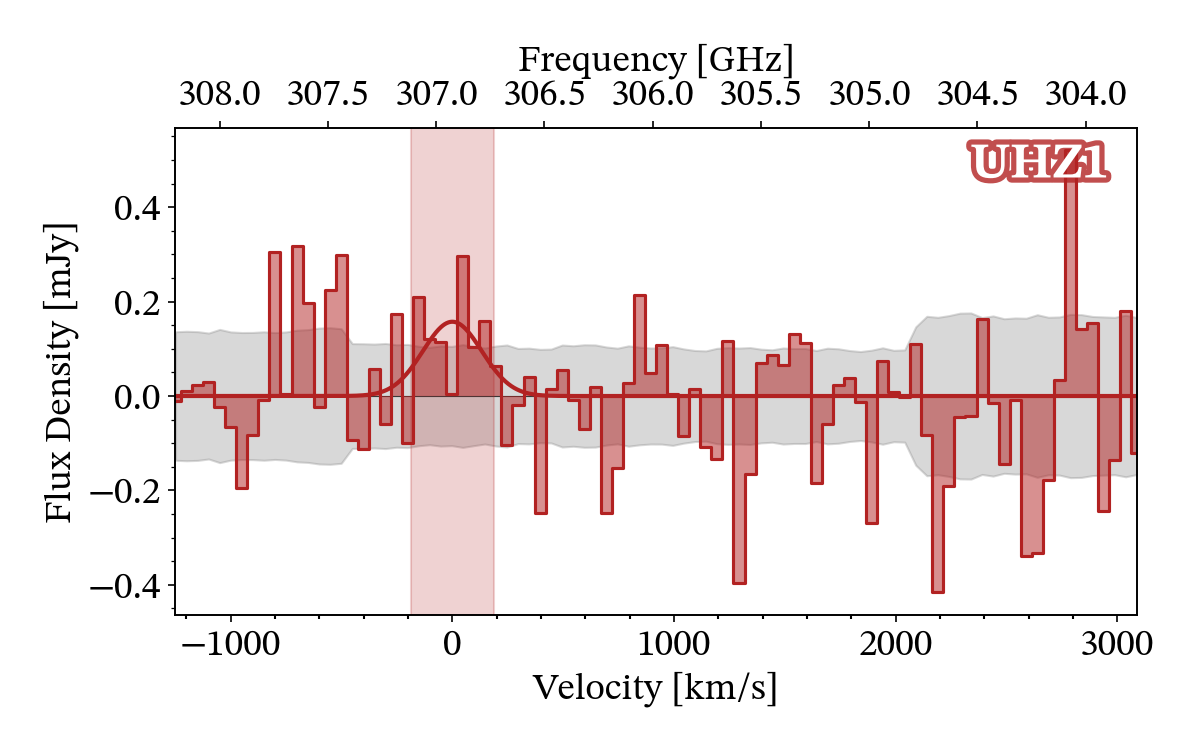}
    \caption{\textbf{Top:} 1D spectrum of UHZ1, extracted in a $0.8''$ aperture centered on its \textit{JWST}/NIRCam position. The expected location of the \oiii{} line based on the fiducial \textit{JWST}/NIRSpec redshift of UHZ1 from \citet[][$z=10.073 \pm 0.002$; G23]{goulding2023} and MIRI H$\alpha$-based redshift from \citet[][$z=10.054 \pm 0.011$; AM26]{alvarez-marquez2026} is indicated through the black errorbars. The red vertical shading corresponds to a width of $\pm100\,\mathrm{km/s}$ around these redshifts. While no emission is seen at the NIRSpec-based redshift, a faint possible signal is seen at the MIRI-based value. \textbf{Bottom:} Peak-pixel spectrum extracted from the faint emission peak identified at the MIRI-based redshift, which is offset by $\sim0.4''$ from the nominal source center of UHZ1 (Figure \ref{fig:continuum_UC10646_tapered}). A Gaussian fit to the emission is overplotted as the solid, red line. Given the low S/N of this feature, we do not consider it a robust \oiiil{} detection, and instead adopt an upper limit on the line flux.}
    \label{fig:spectra1D_UHZ1}
\end{figure}

\begin{figure}
    \centering
    \includegraphics[width=0.75\textwidth]{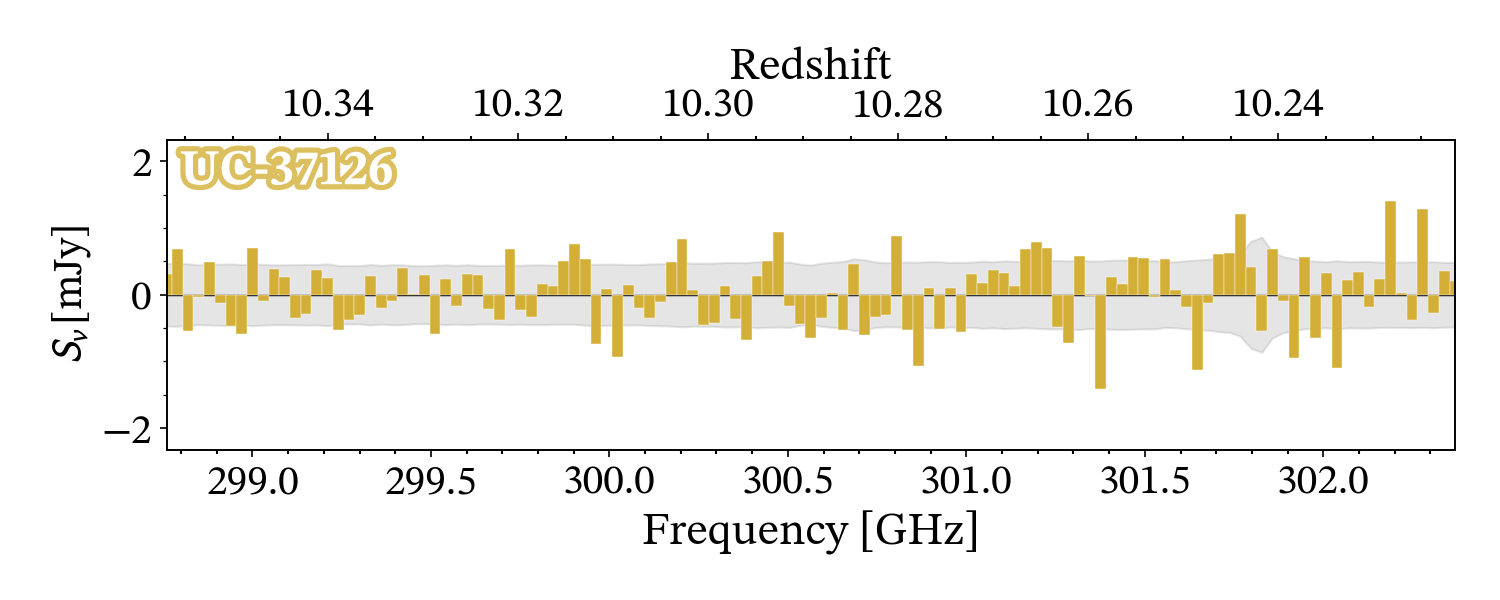}
    \includegraphics[width=0.75\textwidth]{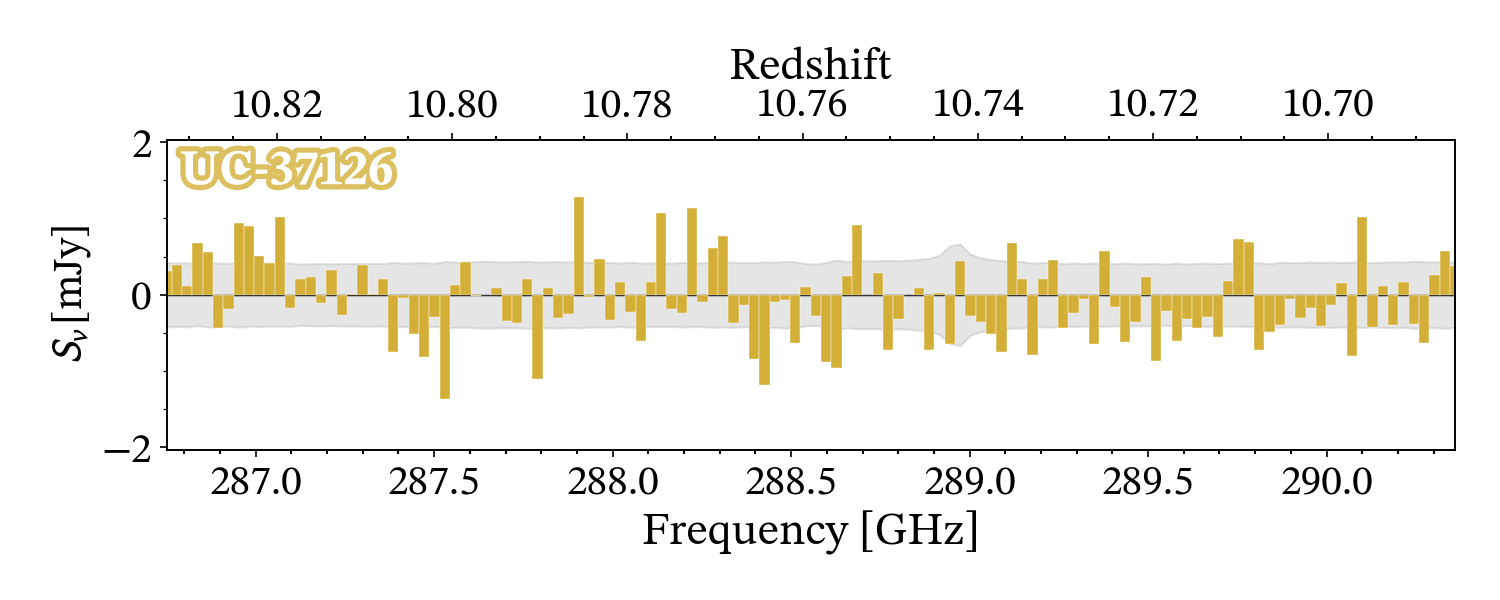}
    \caption{1D spectra of UNCOVER-37126 extracted in a $0.5''$-radius aperture. The redshift of UNCOVER-37126 is uncertain due to a lack of confident emission line detections in both NIRSpec and MIRI, and we therefore show the spectra extracted in both basebands, covering $z=10.22 - 10.36$ and $z=10.68 - 10.83$, respectively (c.f., a fiducial redshift of $z=10.255$). No high-significance feature likely to correspond to the \oiii{} line is present in the ALMA data.}
    \label{fig:spectra1D_UC37126}
\end{figure}

We show the 1D spectra of UHZ1 and UNCOVER-37126 -- for which the \oiii{} line was not detected -- in Figure \ref{fig:spectra1D_UHZ1} and \ref{fig:spectra1D_UC37126}, respectively.

For the former, we see no evidence for any \oiii{} emission at the expected frequency, assuming a fiducial redshift of $z=10.073 \pm 0.002$ as based on the NIRSpec prism spectrum from \citet{goulding2023}. However, we note that, if we were to adopt the (more uncertain) H$\alpha$-based redshift of $z=10.054 \pm 0.011$ based on the recent MIRI/LRS observations of \citet{alvarez-marquez2026}, there is a hint of faint positive emission that could possibly correspond to the \oiii{} line.

Upon closer inspection, the centroid of this feature is offset from the nominal position of UHZ1 by $\sim0.4''$ towards the south-east (right panel of Figure \ref{fig:continuum_UC10646_tapered}). We extract a peak-pixel spectrum from the centroid of the emission feature from a datacube with $50\,\mathrm{km/s}$ channels, and fit it with a Gaussian (bottom panel of Figure \ref{fig:spectra1D_UHZ1}). We note that no good fit is obtained when using our fiducial $0.8''$-radius aperture extraction given the low signal-to-noise of the feature.

The Gaussian fit yields a redshift of $z=10.055 \pm 0.003$, in agreement with the H$\alpha$-based redshift obtained by \citet{alvarez-marquez2026}. The inferred line flux and width are $S_\text{\oiiinowave{}} = 46_{-32}^{+45}\,\mathrm{mJy\,km/s}$ and $\mathrm{FWHM} = 312 \pm 172\,\mathrm{km/s}$, respectively. In the moment-0 map (Figure \ref{fig:continuum_UC10646_tapered}), the feature has a significance of $3.6\sigma$.

We note that the line flux obtained from the Gaussian fit is fully consistent with the $3\sigma$ upper limit quoted in Table \ref{tab:fluxes}, which is (conservatively) based on an aperture rather than a peak-pixel extraction. Given the low-S/N nature of the $z\approx10.055$ feature, as well as the apparent spatial offset with respect to the \textit{JWST}/NIRCam emission, we do not consider the \oiii{} line to be detected in UHZ1, and thus adopt an upper limit on the line flux throughout this work. Deeper ALMA Band 7 observations are needed to confirm whether the observed emission indeed corresponds to \oiii{}.
\\ 

For UNCOVER-37126, we show both redshift intervals covered by the ALMA Band 7 observations in Figure \ref{fig:spectra1D_UC37126}, given that its spectroscopic redshift currently remains uncertain as no emission lines have been confidently detected in its \textit{JWST}/NIRSpec prism spectrum, nor in recent MIRI/LRS observations \citep{fujimoto2024_uncover,marques-chaves2026}. As discussed in Section \ref{sec:observations_targets}, we adopt a fiducial redshift of $z=10.255$ for UNCOVER-37126 based on \citet{fujimoto2024_uncover}. However, Lyman-break-derived spectroscopic redshifts appear to be consistently underestimated compared to emission-line derived ones due to Damped Lyman-$\alpha$ Absorption (DLA), believed to arise from neutral hydrogen along the line of sight \citep[e.g.,][]{deugenio2024,heintz2025_primal,witstok2025}. The net effect is that the Lyman break is both smoothened, and shifted to redder wavelengths, to an extent that depends on the exact neutral hydrogen column density $N_H$. If not accounted for, the Lyman break is thus interpreted to be at longer wavelengths, leading to an overestimated redshift. In our ALMA observations targeting UNCOVER-37126, we did not explicitly account for DLA effects, and instead adopted a standard spectral setup covering the \oiii{} line in one of the two basebands, while maximizing continuum sensitivity in the other. We note, however, that \citet{marques-chaves2026} find UNCOVER-37126 to exhibit a relatively sharp Lyman break, arguing against the presence of a large HI column density. They moreover suggest that the lack of detected emission lines in MIRI are due to a high escape fraction of ionizing photons. In such a scenario, the \oiii{} luminosity would similarly be suppressed.

\section{B.\ Possible extended dust continuum emission in UNCOVER-10646}
\label{app:dust_continuum_UC10646}

\begin{figure}
    \centering
    \includegraphics[width=0.24\linewidth]{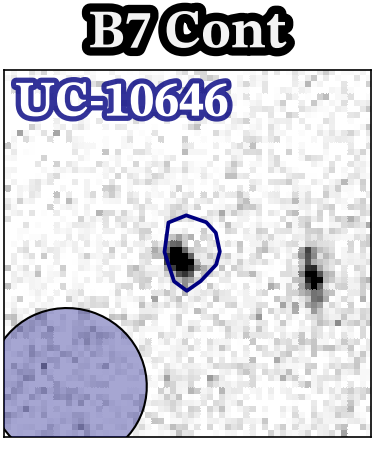}
    \includegraphics[width=0.24\linewidth]{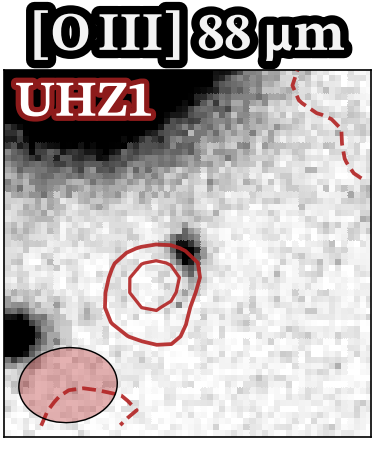}
    \caption{\textbf{Left:} ALMA Band 7 dust continuum image of UNCOVER-10646 ($3''\times3''$), tapered to a $\sim1.3''$ resolution, overplotted on NIRCam F150W imaging. The tapered map reveals a $\sim2\sigma$ dust continuum signal co-spatial with the NIRCam emission of the $z=8.51$ galaxy. Given that UNCOVER-10646 appears to host a powerful outflow (Section \ref{sec:discussion_outflow}), its dust may be also efficiently expelled from the galaxy resulting in an extended dust reservoir that is recoverable only in the tapered imaging. However, given the low S/N of the emission, we regard the possible dust continuum detection as tentative. \textbf{Right:} Moment-0 map ($3''\times3''$) of the tentative emission feature detected for UHZ1, which may correspond to the \oiii{} line at $z=10.055 \pm 0.003$. The nominal significance of the feature, which is offset from the NIRCam emission (F444W, background) of the galaxy by $\sim0.4''$, is $3.6\sigma$ (contours show $\pm2, 3\sigma$). We do not consider this a robust detection, and instead adopt an upper limit on the \oiii{} line flux throughout this work.}
    \label{fig:continuum_UC10646_tapered}
\end{figure}

With the fiducial natural weighting, our continuum observations of UNCOVER-10646 attain a resolution of $\sim0.6''$ (Table \ref{tab:observations}). While this exceeds the rest-UV and -optical size of the system as seen with NIRCam (Weaver et al.\ in preparation), it has been suggested that dust in particularly luminous galaxies can extend out to larger radii due to radiation-driven outflows \citep{ferrara2023,ziparo2023}. To investigate the possibility of spatially-extended dust emission in our four $z>8.5$ targets, we consider their $>1.0''$ tapered continuum maps. These reveal a $\sim2\sigma$ dust continuum feature only in UNCOVER-10646, fully co-spatial with the NIRCam emission (Figure \ref{fig:continuum_UC10646_tapered}). The peak flux density of this feature equals $S_\mathrm{cont} = 48.4 \pm 22.0\,\mu\mathrm{Jy/beam}$, and is marginally consistent with the upper limit of $S_\mathrm{cont} < 43.3\,\mu\mathrm{Jy}$ provided in Table \ref{tab:fluxes}, which assumes unresolved emission. With our fiducial dust temperature of $T_\mathrm{dust} = 50\,\mathrm{K}$ (Section \ref{sec:discussion_continuum}), taking this signal at face value would imply UNCOVER-10646 contains a dust mass of $\log(M_\mathrm{dust}/M_\odot) \approx 6.3$ and has an obscured star formation rate of $\mathrm{SFR}_\mathrm{IR} \approx 9.4\,M_\odot\,\mathrm{yr}^{-1}$. However, given the low S/N of the continuum signal, we regard it as tentative and subject to further confirmation.

\section{C.\ The \oiii{}-SFR relation with homogenized star formation rate measurements}
\label{app:oiii_SFR}

In Section \ref{sec:discussion_oiii_sfr}, we present the \oiii{}-SFR relation of our sample (see also Figure \ref{fig:oiiiSFR}). However, as discussed in Section \ref{sec:observations_targets}, our star formation rate measurements are drawn from the literature, and were therefore obtained under a slightly different set of assumptions for our individual targets. In an attempt to homogenize these measurements, we consider two limiting cases that should bracket the true SFRs of our targets. As a lower limit, we adopt their UV SFRs, uncorrected for obscuration, while as an upper limit we adopt total SFRs as the sum of the UV and (undetected) IR emission. Both UV- and IR trace star formation on similar timescales of $\lesssim100\,\mathrm{Myr}$ \citep{kennicutt2012}, or possibly even shorter \citep[$\sim30\,\mathrm{Myr}$, e.g.,][]{fisher2025b}. Moreover, the local comparison sample from \citet{delooze2014} also adopts UV + IR-based SFRs, using GALEX FUV and \textit{Spitzer}/MIPS $24\,\mu\mathrm{m}$ data, respectively. The derivation of upper limits on the IR SFRs of our targets is discussed in Section \ref{sec:discussion_continuum}. We note that these limits depend on the assumed dust temperature, for which we adopted a fiducial $T_\mathrm{dust}=50\,\mathrm{K}$. A higher value of, for example, $70\,\mathrm{K}$ would increase the upper limits on $L_\mathrm{IR}$ and thus on the obscured SFRs by $\approx0.45\,\mathrm{dex}$.

For three of our targets -- DHZ1, UHZ1 and UNCOVER-37126 -- the \oiii{}-SFR relation does not significantly change when adopting these UV- and UV+IR-based values. For UNCOVER-10646, as already discussed in Section \ref{sec:discussion_continuum}, the upper limit inferred from its UV+IR-based SFR is lower than that obtained from SED fitting and consequently it falls marginally above the local \oiii{}-SFR relation for dwarf galaxies from \citet{delooze2014}. We note that, if we consider only the narrow \oiii{} component of UNCOVER-10646, which should trace just its ISM rather than also the outflow component, we do find it to fall onto the local dwarf galaxy relation.

\begin{figure}
    \centering
    \includegraphics[width=0.7\linewidth]{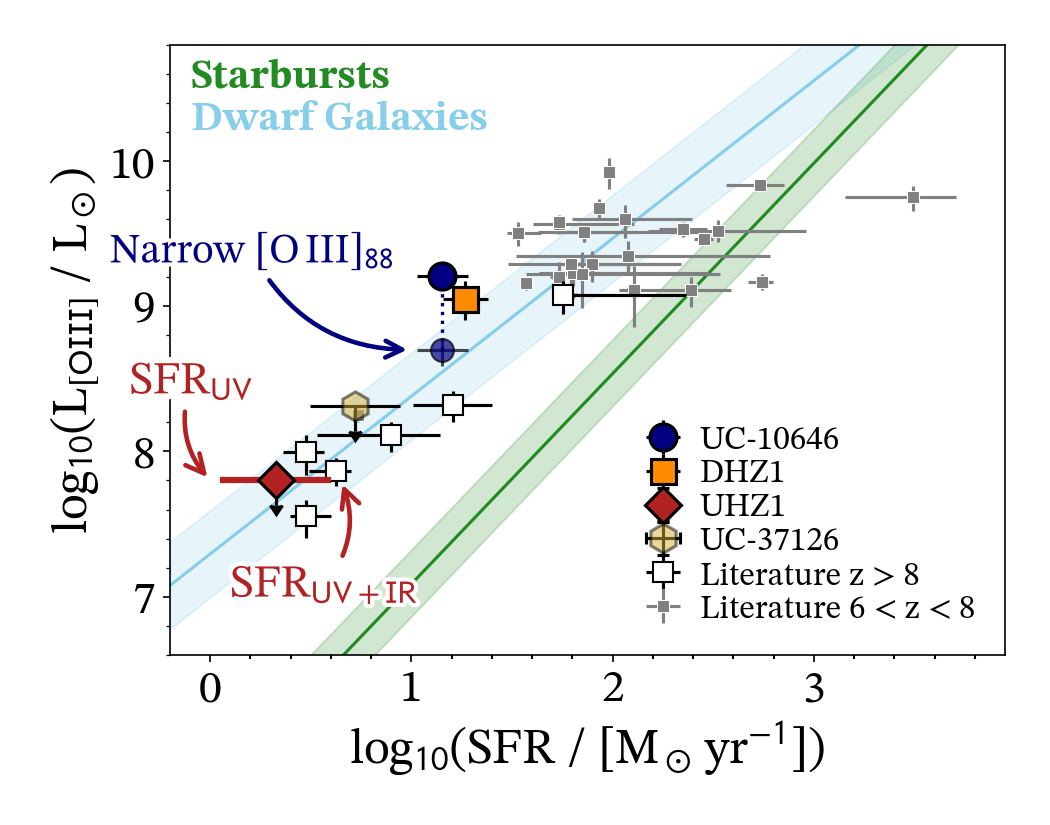}
    \caption{An alternative \oiii{}-SFR relation for our sample. Instead of showing the SFRs obtained from SED fitting as in Figure \ref{fig:oiiiSFR}, we now show the UV- and UV+IR SFRs, which should be regarded as lower and upper limits, respectively, given that no dust continuum emission is robustly detected from our targets. The horizontal errorbars on the datapoints bracket this range of SFRs, with the lower (upper) value representing the UV only (UV + IR) star formation rate. For UNCOVER-10646, we also show the line luminosity considering only the narrow component of the \oiii{} line as the smaller blue marker. Other symbols are as in Figure \ref{fig:oiiiSFR}.}
    \label{fig:oiiiSFR_UV_IR}
\end{figure}

\section{D.\ Further details on the possible ionized outflow in UNCOVER-10646}
\label{app:outflow_UC10646}

\subsection{Briggs-weighted imaging of the \oiii{} line}

To investigate whether the \oiii{} emission from UNCOVER-10646 can be spatially resolved, we re-image the line at a higher angular resolution, adopting Briggs weighting \citep{briggs1995} with a robust parameter of $0.5$. This yields a datacube with a beam size of $0.48''\times0.39''$ at the central frequency of the \oiii{} line -- or about $1.4\times$ better angular resolution than the naturally-weighted cube -- at the expense of a $\sim15\%$ penalty in sensitivity. We have additionally attempted more aggressive Briggs weighting schemes (robust $\in[0.0, -0.5]$) yielding beam sizes down to $0.37'' \times 0.31''$, but found no differences with the results detailed below. Given the reduction in S/N at these higher-resolution weightings, we focus below on the $\mathrm{robust} = 0.5$ datacube.

\begin{figure}
    \centering
    \includegraphics[width=0.51\linewidth]{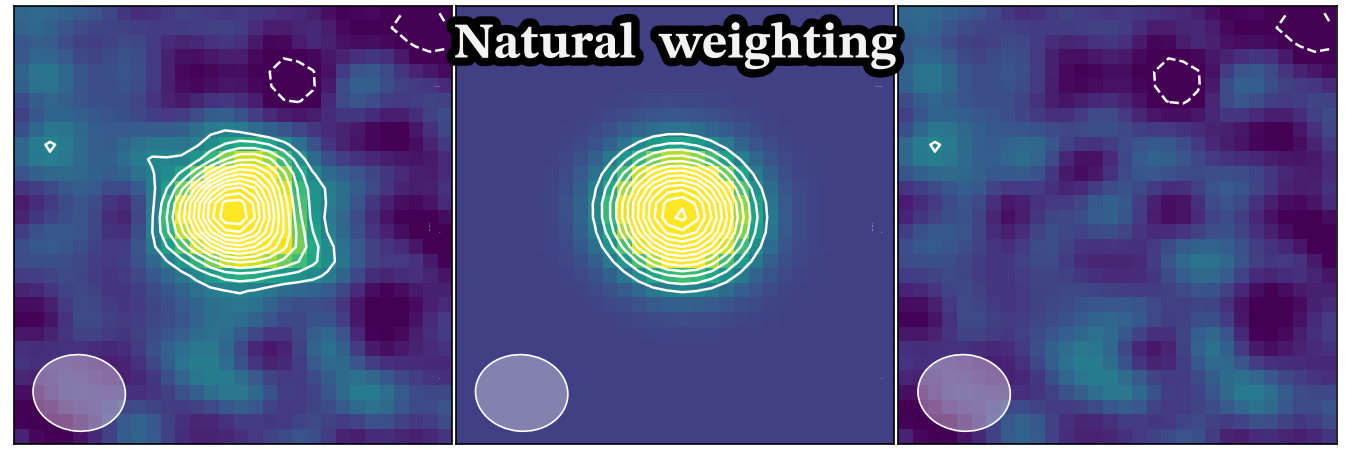}
    \includegraphics[width=0.51\linewidth]{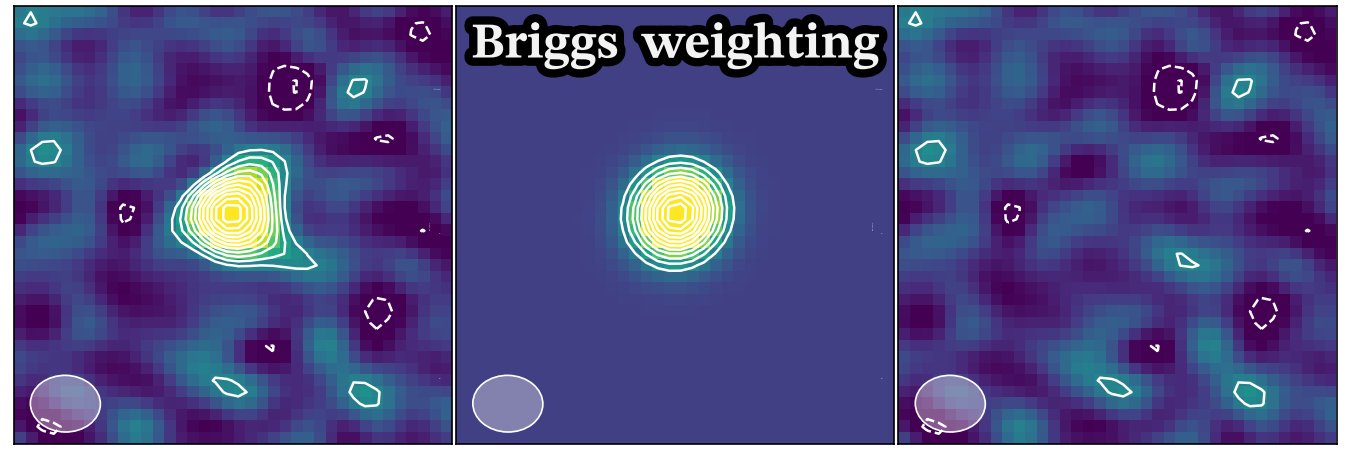}
    \caption{Moment-0 maps (\textit{left}), {\sc{CASA imfit}} Gaussian models (\textit{middle}) and residuals (\textit{right}) of the naturally-weighted (\textit{top}) and Briggs-weighted (\textit{bottom}) \oiii{} emission of UNCOVER-10646. The cutouts span a $3''\times3''$ area and contours show emission at the $\pm2,3,4,\ldots\times\sigma$ level. The \oiii{} line is not confidently resolved in the natural- and Briggs-weighted maps, suggesting the putative outflow in the 1D spectrum of UNCOVER-10646 (Figure \ref{fig:singleDualGaussian}) is likely to be compact, or possibly directed along the line of sight towards us within a narrow opening angle.}
    \label{fig:UC10646_imfit}
\end{figure}

We show the Briggs-weighted moment-0 map of the \oiii{} emission -- extracted in the same manner as in Section \ref{sec:observations_fluxMeasurements}, albeit now in a smaller $0.6''$ aperture given the higher angular resolution -- as well as a 2D Gaussian fit to the moment-0 map with {\sc{CASA}} {\sc{imfit}} in the bottom row of Figure \ref{fig:UC10646_imfit}. The best fit corresponds to an approximately circular Gaussian of $(0.49'' \pm 0.04'') \times (0.47'' \pm 0.04'')$, suggesting UNCOVER-10646 may be marginally resolved along the beam minor axis, but not along the beam major axis. Indeed, from the fit we infer a maximum beam-deconvolved size for its \oiii{} emission of $(< 0.36'') \times (< 0.22'')$.

For completeness, we show a similar fit to the naturally-weighted moment-0 map in the upper row of Figure \ref{fig:UC10646_imfit}. This yields a size of $(0.74'' \pm 0.06'') \times (0.65'' \pm 0.05'')$, which corresponds to a beam-deconvolved extent of $(0.37'' \pm 0.14'') \times (0.31'' \pm 0.21'')$. This too suggests that UNCOVER-10646 is, at best, marginally resolved along the beam major axis. However, we deem the upper limit obtained from the Briggs-weighted map to be more robust, given its higher angular resolution, and therefore adopt a fiducial size of the UNCOVER-10646 system of $0.30'' \pm 0.10''$ in Section \ref{sec:discussion_outflow}.

\subsection{Peak-pixel and aperture extractions in the dirty and clean datacubes}

\begin{figure}
    \centering
    \includegraphics[width=0.49\textwidth]{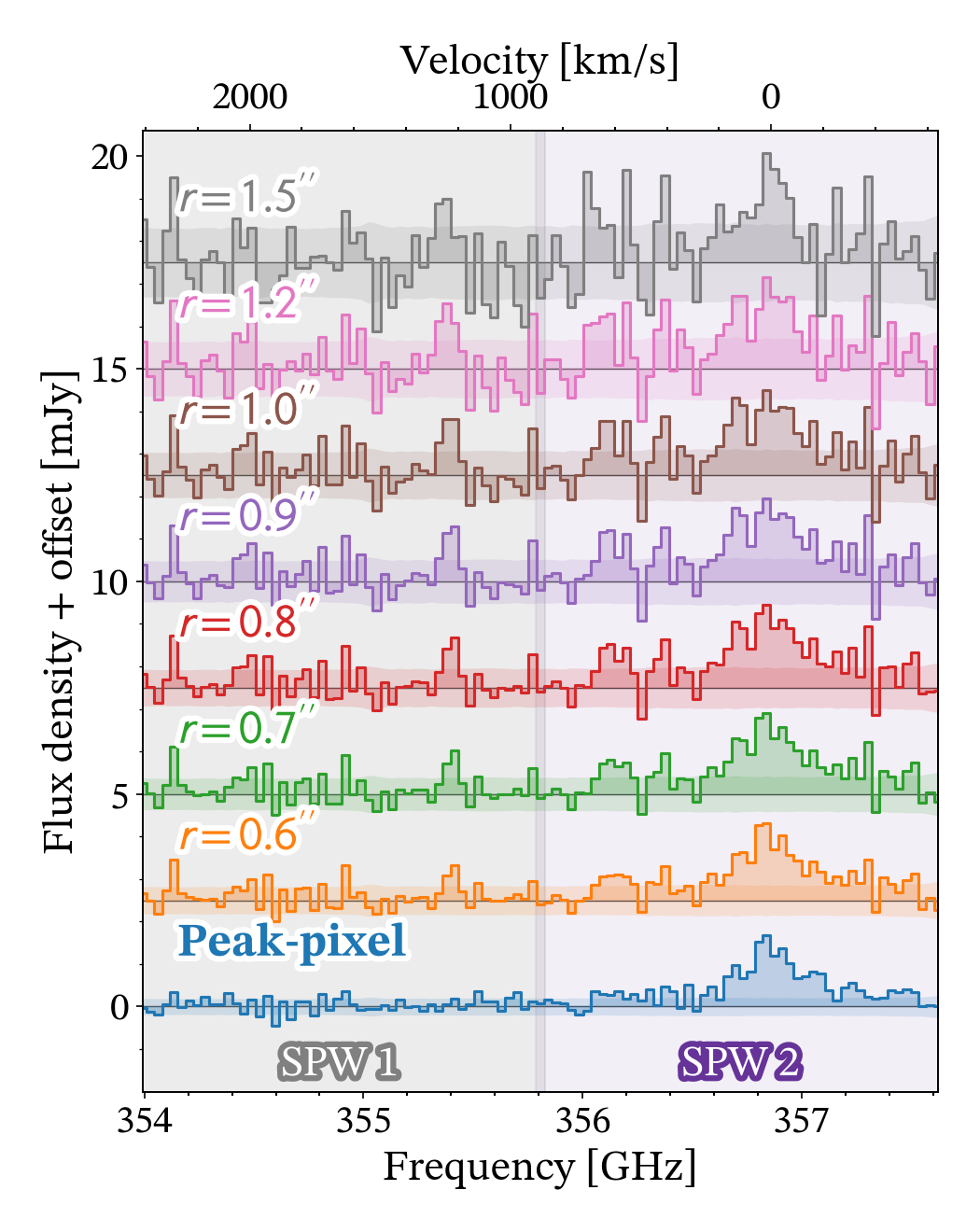}
    \includegraphics[width=0.49\textwidth]{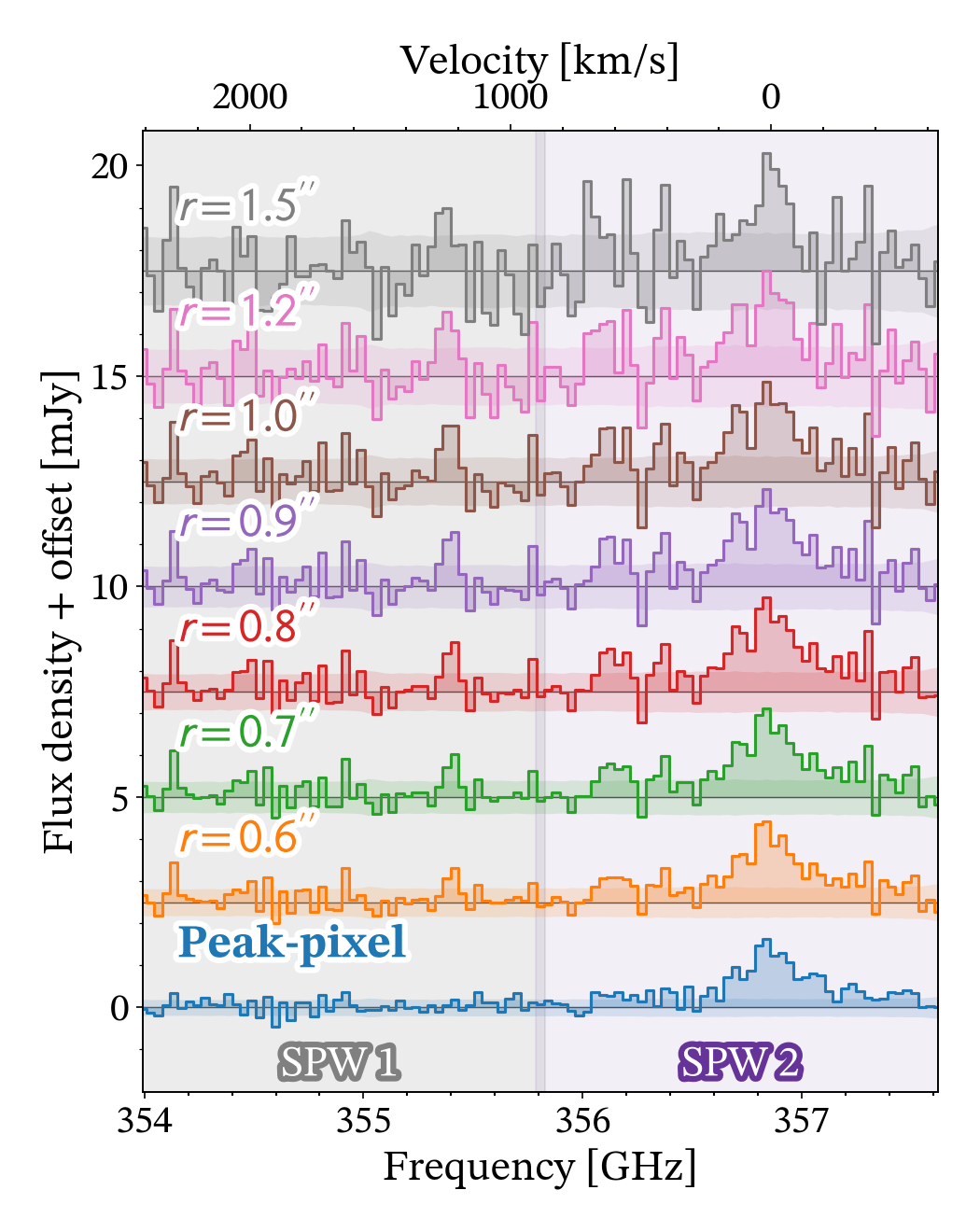}
    \caption{Peak-pixel (blue) and aperture spectra of UNCOVER-10646, extracted from the {\sc{clean}}ed (\textit{left}) and dirty datacubes (\textit{right}). Spectra are offset in steps of $2.5\,\mathrm{mJy}$ for visual clarity. The frequency ranges of the two concatenated spectral windows in the ALMA baseband containing the emission line are shown through the background grey and purple shading, respectively. High-velocity \oiii{} emission is clearly detected in all extractions, although the noise level rapidly increases when large aperture radii are adopted ($r\gtrsim1.0''$). }
    \label{fig:comparisonPeakApertureSpectra}
\end{figure}

As discussed above, we find that the outflow in UNCOVER-10646 cannot be reliably resolved even in Briggs-weighted imaging, and we therefore optimize its S/N by extracting a peak-pixel spectrum in Section \ref{sec:discussion_outflow}. We compare the peak-pixel spectrum to a set of aperture extractions using a range of radii $r = 0.6'' - 1.5''$ in Figure \ref{fig:comparisonPeakApertureSpectra} for both the {\sc{clean}} (left panel) and dirty (right) cubes. As extensively discussed in \citet{meyer2022}, faint emission in the wings of an emission line is generally not {\sc{clean}}ed, unlike the brighter channels covering the line center. This could result in systematic differences between the {\sc{clean}} and dirty fluxes. However, the right panel of Figure \ref{fig:comparisonPeakApertureSpectra} shows that the 1D spectra extracted from the dirty map also contain positive emission in the wings of the \oiii{} line, suggesting that the {\sc{clean}} algorithm does not induce spurious flux in the emission line wings. We note that the {\sc{clean}} and dirty spectra are very similar overall, and we have moreover confirmed that the high-velocity emission remains when performing a more aggressive interactive {\sc{clean}} run on the visibility data, compared to the fiducial auto-masked {\sc{clean}} performed in Section \ref{sec:observations}.

Furthermore, high-velocity \oiii{} emission is clearly visible in all aperture extractions in Figure \ref{fig:comparisonPeakApertureSpectra}, although the signal-to-noise ratio of emission in the wings decreases for apertures well beyond the beam size. This is not surprising, given that UNCOVER-10646 is not confidently resolved at the current resolution ($\sim0.6''$, using the naturally-weighted datacube), and the likely compact nature of the outflow. Finally, while the high-velocity emission appears contained within a single spectral window (SPW 2 in Figure \ref{fig:comparisonPeakApertureSpectra}), the positive signal does not abruptly vanish at the SPW boundary, which would otherwise have been suspicious. \\

Given that UNCOVER-10646 appears marginally resolved along the beam minor axis (see the above discussion), we quantify whether any flux might be lost in a peak-pixel extraction, compared to the fiducial aperture measurements. We do so in two different ways: first, we compare our fiducial flux measurements obtained from a single-Gaussian fit to the aperture spectrum, and a dual-Gaussian fit to the peak-pixel spectrum. These yield values of $S_\text{\oiii{}}^\mathrm{aper} = 824_{-145}^{+159}\,\mathrm{mJy\,km/s}$ and $S_\text{\oiii{}}^\mathrm{peak-pxl} = 752_{-81}^{+93}\,\mathrm{mJy\,km/s}$, respectively (Table \ref{tab:fluxes}), which are fully consistent within the uncertainties, and thus suggest that no significant flux is lost when adopting a peak-pixel spectrum.

Secondly, we measure the peak and aperture fluxes directly from the moment-0 maps presented in Figure \ref{fig:UC10646_imfit}. This analysis reveals that for the Briggs-weighted (naturally-weighted) moment-0 map, an aperture extraction yields a higher line flux by a factor of $1.25 \pm 0.18$ ($1.32 \pm 0.15$) compared to the peak-pixel flux. If these slightly larger aperture fluxes are due to \oiii{} emission from the galaxy-wide ISM of UNCOVER-10646 (i.e., the narrow component), rather than a compact outflow, our outflow rates quoted in Section \ref{sec:discussion_outflow} are not affected, although in this case the total ionized ISM mass of the system may be underestimated by at most $\sim30\,\%$. Between these two methods, however, we find no strong evidence of a significant loss of line flux in the peak-pixel extraction, suggesting this unresolved analysis is appropriate.

Overall, the above analysis suggests the observed high-velocity emission is not an artifact of the imaging procedure, or the extraction method. Still, we stress that higher resolution \oiii{} data are needed to confirm the putative outflow.

\subsection{Fitting two Gaussians to the fiducial aperture spectrum}

Based on Bayesian model comparison, we find that a double-Gaussian fit is statistically preferred over a single Gaussian component for the peak-pixel extraction (Figure \ref{fig:singleDualGaussian}). However, as can be seen in Figure \ref{fig:comparisonPeakApertureSpectra}, the broad component appears less clear in aperture-extracted 1D spectra.

\begin{figure}
    \centering
    \includegraphics[width=0.85\linewidth]{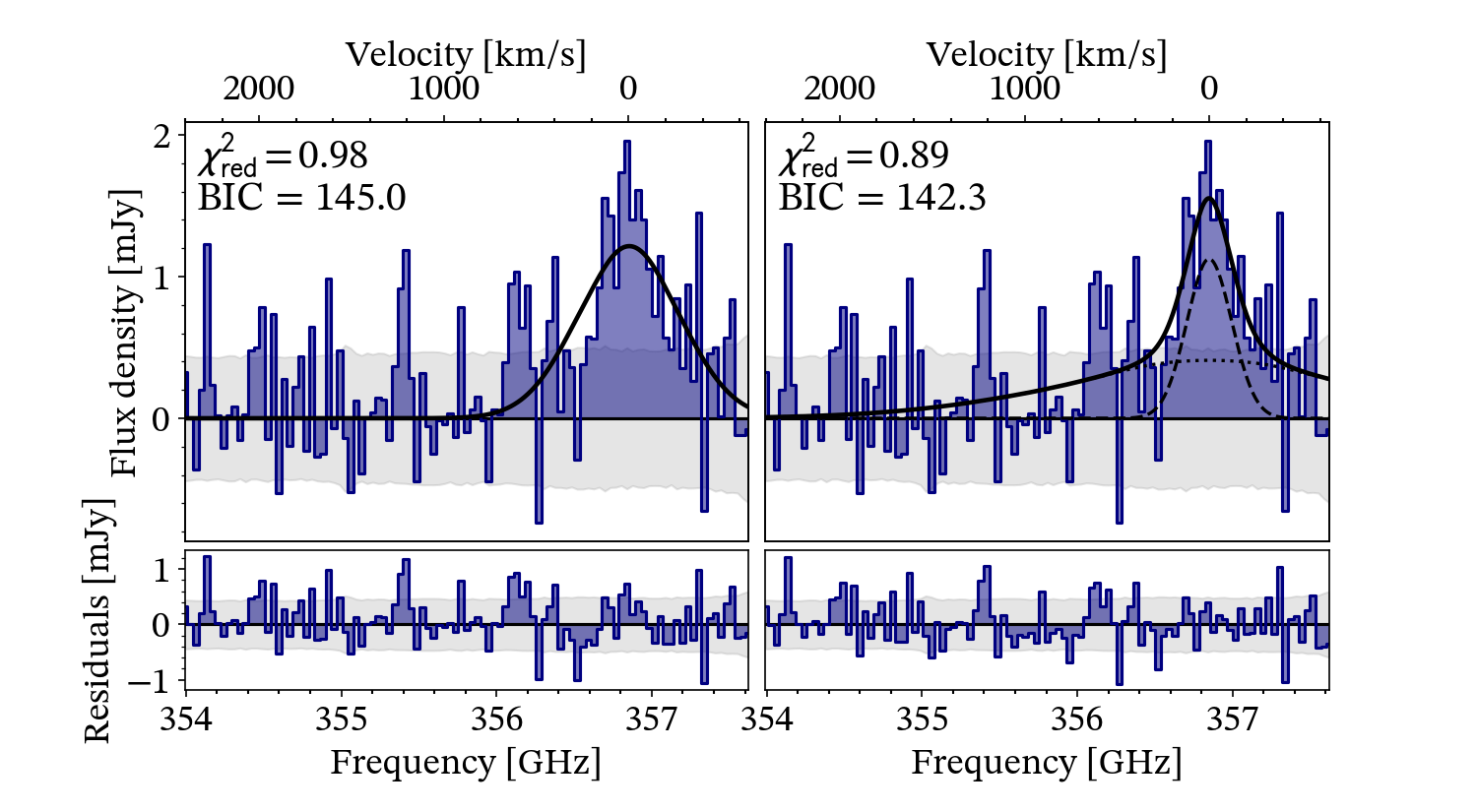}
    \caption{Single (\textit{left}) and dual (\textit{right}) Gaussian fits to the fiducial aperture-extracted spectrum of UNCOVER-10646 (Figure \ref{fig:spectra1D}). Unlike is the case for the peak-pixel spectrum (Figure \ref{fig:singleDualGaussian}), the aperture spectra show only marginal evidence that a broad Gaussian component is required, with a minor improvement in reduced-$\chi^2$ and $\Delta\mathrm{BIC} = 2.7$. This suggests that, if the outflow is real, it is likely to be compact.}
    \label{fig:singleDualGaussianApertureSpectra}
\end{figure}

We quantify this in Figure \ref{fig:singleDualGaussianApertureSpectra}, where we fit the fiducial 1D spectrum of UNCOVER-10646 -- extracted in a $0.8''$-radius aperture, as in Figure \ref{fig:spectra1D} -- with both a single and two Gaussian components. We find the dual Gaussian is marginally preferred based on a simple $\chi^2_\mathrm{red}$ comparison. The $\Delta\mathrm{BIC} = 2.7$ similarly implies `positive evidence' of the dual-Gaussian model being preferred as per the \citet{raftery1995} scale, albeit at lower statistical significance than for the peak-pixel spectrum. This suggests that, if the outflow is real, it must be compact for its flux to be diluted in aperture-based extractions.

\subsection{Channel maps}

\begin{figure}
    \centering
    \includegraphics[width=0.95\linewidth]{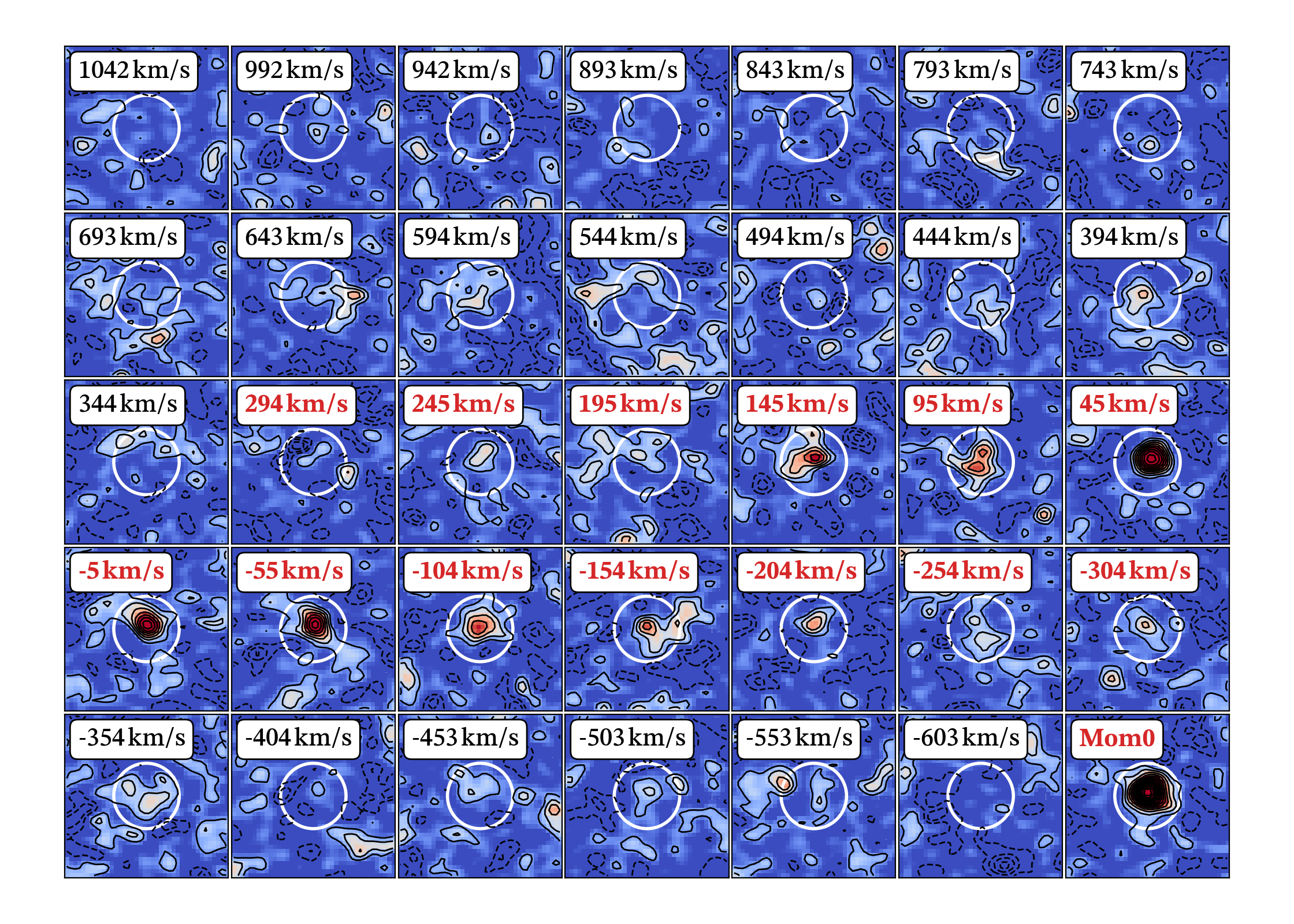}
    \caption{Channel maps for the naturally-weighted, {\sc{clean}} datacube of UNCOVER-10646, spanning a broad velocity range across its narrow and putative broad \oiii{} emission in $50\,\mathrm{km/s}$ channels. Cutouts are $4''\times4''$, and our fiducial $0.8''$-radius extraction aperture is overplotted in white. The colorscale runs linearly between $0-5\sigma$ where $\sigma$ is the RMS in a given channel. The contours start at $\pm1\sigma$ and increase in steps of $\pm1\sigma$ to highlight low-level emission. Negative contours are dashed. The moment-0 map, collapsed across the channels marked in red, is shown in the lower right corner of the plot.}
    \label{fig:channelMaps}
\end{figure}

We show channel maps for UNCOVER-10646, ranging from $-600$ to $+1050\,\mathrm{km/s}$ around the central frequency of its \oiii{} line, in Figure \ref{fig:channelMaps}.

\end{document}